\documentclass{emulateapj}

\newcommand{\kms}{km~s$^{-1}$\,}
\newcommand{\mass}{{\cal M}}

\newcommand{\msun}{{\mass_\odot}}

\newcommand{\beq}{\begin{equation}}
\newcommand{\eeq}{\end{equation}}
\newcommand{\mh}{M_{\bullet}}
\newcommand{\msig}{$\mh-\sigma$}

\newcommand{\errNol}{\Delta(N^{\rm obs}_l(v_y^\prime))}

\shorttitle{3-Integral Models}
\shortauthors{Valluri, Merritt \& Emsellem }

\begin{document}

\title{Difficulties with Recovering The Masses of Supermassive Black Holes 
from Stellar Kinematical Data}
\bigskip\bigskip

\author{ Monica Valluri}
\medskip
\affil{Department of Astronomy and Astrophysics \\
and Center for Cosmological Physics\\ University of Chicago\\
5640 S. Ellis Avenue, Chicago 60637}
\email{valluri@oddjob.uchicago.edu}
\bigskip
\author{David Merritt}
\medskip
\affil{Rutgers University\\
Department of Physics and Astronomy\\
New Brunswick, NJ 08903}
\email{merritt@physics.rutgers.edu}
\author{Eric Emsellem}
\medskip
\affil{Centre de Recherche Astronomique de Lyon\\
9 av. Charles Andr\'e, 69561 Saint-Genis Laval Cedex\\
France}
\email{emsellem@obs.univ-lyon1.fr}
\bigskip

\slugcomment{Accepted for publication in the Astrophysical Journal}
\clearpage
\newpage

\begin{abstract}
We investigate the ability of three-integral, axisymmetric,
orbit-based modeling algorithms to recover the parameters defining the
gravitational potential (mass-to-light ratio $\Upsilon$ and black hole
mass $\mh$) in spheroidal stellar systems using stellar kinematical
data.  We show that the potential estimation problem is generically
under-determined when applied to long-slit kinematical data of the kind used
for most black hole mass determinations to date. A range of parameters
($\Upsilon,\mh$) can provide equally good fits to the data, making it
impossible to assign best-fit values.  The indeterminacy arises from
the large variety of orbital solutions that are consistent with a
given mass model.  We demonstrate the indeterminacy using a variety of
data sets derived from realistic models as well as published
observations of the galaxy M32.  The indeterminacy becomes apparent
only when a sufficiently large number of distinct orbits are supplied
to the modeling algorithm; if too few orbits are used, spurious minima
appear in the $\chi^2(\Upsilon,\mh)$ contours, and these minima do not
necessarily coincide with the parameters defining the gravitational
potential.

We show that the range of degeneracy in $\mh$ depends on the degree to
which the data resolve the radius of influence $r_h$ of the black
hole.  For ${\rm FWHM}/2r_h\gtrsim 0.5$, where FWHM refers to the
instrumental resolution, we find that only very weak constraints can
be placed on $\mh$. In the case of M32, our reanalysis demonstrates
that when a large orbit library is used, data published prior to 2000
(${\rm FWHM}/2r_h\approx 0.25$) are equally consistent with black hole
masses in the range $1.5\times10^6\msun<\mh<5\times 10^6\msun$, with
no preferred value in that range. Exactly the same data can reproduce
previous published results with smaller orbit libraries. While the
HST/STIS data for this galaxy (${\rm FWHM}/2r_h\approx 0.06$) may
overcome the degeneracy in $\mh$, HST data for most galaxies do not
resolve the black hole's sphere of influence and in these galaxies the
degree of degeneracy allowed by the data may be greater than
previously believed.

We investigate the effect of regularization, or smoothness
constraints, on the degree of degeneracy of the solutions.  Enforcing
smoothness reduces the range of acceptable models, but we find no
indication that the true potential can be recovered simply by
enforcing smoothing.  For a given smoothing level, all solutions in
the minimum-$\chi^2$ valley exhibit similar levels of noise; as the
smoothing is increased, there is a systematic shift in the midpoint of
the $\chi^2$ valley, until at a high level of smoothing the solution
is biased with respect to the true solution. These experiments
suggest both that the indeterminacy is real -- i.e., that it is not an
artifact associated with non-smooth solutions -- and that there is no
obvious way to choose the smoothing parameter to ensure that
the correct solution is selected.

\end{abstract}


\bigskip

\section{Introduction}
\label{sec:intro}

Supermassive black holes (SBHs) are believed to be the central engines
of active galactic nuclei and quasars \cite{Lynden-Bell69}.  A
substantial fraction of the mass involved in the energy production is
expected to collapse onto the central black hole.  There is now
irrefutable dynamical evidence for a SBH at the center of our Galaxy
\cite{Genzel97, Ghez98} and in NGC 4258 \cite{Miyoshi95}.  In addition
there is compelling evidence that compact mass concentrations --
probably SBHs -- exist in the nuclei of a handful of other galaxies.
The STIS GTO program (\cite{Joseph01, Bower01}; Merritt et al. 2001), and
several HST GO projects \cite{Sarzi01, Barth01, Hughes01, Gebhardt02}
have begun to extend the search for SBHs to a sample of roughly a
hundred galaxies.

Before this search was fully underway, a tight empirical correlation
was discovered between the masses of SBHs and the velocities of stars
in their host bulges.  
The \msig\ relation:
\begin{eqnarray}
\mh = (1.48\pm 0.24)\times 10^8 \msun \left({\sigma\over 200 {\rm km\ s}^{-1}}\right)^\alpha, \\
\alpha = 4.65 \pm 0.48\nonumber
\label{eq:ms}
\end{eqnarray}

\cite{FerrareseM00, Gebhardt00b}, relates $\mh$ to a measure of the
stellar velocity dispersion in a region larger than the region
directly influenced by the SBH, $r_h\equiv G\mh/\sigma^2$ (slope and
normalization taken from \cite{MerrittF01b}).  The tightness of the
relation depends crucially on the sample used to define it: SBHs whose
masses were derived from data that resolve $r_h$ define a relation
with negligible scatter, $\tilde\chi^2\lesssim 1$, while including all
published detections regardless of their quality yields a weaker
relation and a different slope \cite{FerrareseM00, MerrittF01a,
MerrittF01b}.  Almost all SBH masses derived from ground-based,
stellar-kinematical data \cite{Magorrian98} scatter above the \msig\
relation defined by the more secure masses.

If the current normalization of the \msig\ relation, equation
(\ref{eq:ms}), is correct, SBHs in the more distant of the Magorrian
et al. (1998) galaxies are too small for their radii of influence to
have been resolved by existing telescopes.  Modeling of such data is
prone to systematic errors, the sign and magnitude of which depend on
the form of the dynamical model fit to the stellar motions and the degree
of under-resolution.  van der Marel (1999) argued that the
two-integral (2I) axisymmetric models used by Magorrian et al. (1998)
were likely to give spuriously large values of $\mh$.

The discovery that the ground-based mass estimates were systematically
high resolved the discrepancies between the mean SBH mass inferred
from quasar statistics and reverberation mapping of (mostly) distant
galaxies, on the one hand, and from the kinematics of nearby galaxies
on the other \cite{Richstone98}.  All techniques now yield a mean
ratio of SBH mass to bulge mass of $\sim 10^{-3}$ and a mean SBH mass
density of $\sim 3\times 10^5\msun {\rm pc}^{-3}$
\cite{Ferrarese01,Tremaine02}.

The biases associated with 2I modeling can in principle be removed if
the data are compared with fully general, ``three integral'' (3I)
axisymmetric models, in which any distribution of orbits is allowed
\footnote{We adopt the standard name for these algorithms even though
orbits in axisymmetric potentials are sometimes characterized by fewer
than three integrals.}.  Such models have been used to estimate $\mh$
in a number of galaxies (\cite{vanderMarel98}; \cite{CrettonB99};
(Emsellem et al. 1999); \cite{Gebhardt00a}; \cite{Gebhardt02};
\cite{Cappellari02};\\ \cite{Verolme02}) .  In contrast to 2I models, 3I
models can precisely reproduce a given mass distribution with many
different orbital populations.  This extra freedom is so great that
one does not necessarily expect to find a unique potential that yields
a best fit to the data; indeed there may exist many choices for the
parameters ($\Upsilon, \mh$) that reproduce the data equally well.
This indeterminacy of potential estimation is well documented
\cite{Merritt93a,Gerhard98}.

In this paper, we discuss the importance of the degeneracy in the
context of stellar-dynamical estimates of $\mh$ in galactic nuclei.
We apply a state-of-the-art 3I modeling algorithm to various data
sets, including previously-analyzed data from M32, as well as
simulated data generated from an axisymmetric model of M32.  We
investigate how accurately a 3I modeling algorithm can recover the
true values of $\mh$ and $\Upsilon$, and how sensitively the estimates
of those quantities, and their errors, depend on the quality of the
data, the character of the data, and the number of orbits included in
the model, and the degree of smoothing applied.

Our conclusion is that the indeterminacy problem is often severe.
Even when modeling ``good'' data, the range of values of $\mh$ that
can reproduce the data equally well is typically very large.  (We
define ``good'' data as data that resolve the SBH's sphere of
influence; extend far enough in radius to constrain the global
mass-to-light ratio; include high-order moments of the line-of-sight
velocity distributions; and have small errors.)  This degeneracy can
formally be reduced by placing restrictions on the allowed functional
form of the stellar distribution function; indeed it was in just this
way that Magorrian et al. (1998) achieved their fits, by restricting
$f$ to a two-integral form.  Another such restriction, common in the
more recent studies, is to force the 3I $f$ to be smooth
\cite{Cretton99, Gebhardt02, Cappellari02, Verolme02}.  Smoothness
constraints might reasonably be expected to guide the optimization
routine away from solutions that vary strongly between the data
points, achieving good fits only by virtue of the discrete character
of the data \cite{MerrittF96,Jalali02}.  However we find no indication
that smoothness on its own can overcome the inherent degeneracy in the
potential estimation problem. Furthermore, if used carelessly,
smoothness constraints can bias the solution, yielding an apparent
best-fit value for $\mh$ which lies far from the true value.

In \S~2 we give a detailed description of our 3I modeling algorithm.
\S~3 reviews the reasons why the potential estimation problem is
expected to be under-determined in the axisymmetric geometry. \S~4
presents a 2I model of M32 that we use as a test case for our
algorithm. \S~5 and \S~6 present detailed results of fits of simulated
data sets derived from the M32 model, and \S~7 describes the results
of a re-analysis of published data for this galaxy. \S~8 describes how
the introduction of additional regularization or smoothness
constraints affects the results when applied to the simulated M32
data.  \S~9 is a discussion of the implications of our results for the
recovery of $\mh$ in nearby galaxies, and \S~10 sums up.

\bigskip

\section{Modeling Algorithm}
\label{sec:algorithm}

\subsection{Density and Potential}
\label{sec:dens_pot}

We construct dynamical models of axisymmetric stellar systems with
mass density $\rho(\varpi,z)$ and potential $\Phi(\varpi,z)$; $z=0$
defines the equatorial plane.  The mass density may contain
contributions from stars, $\rho_*$, as well as other components such
as dark matter or a central black hole.  The contribution to the mass
density from the stars is derived from the luminosity density
$j_*(\varpi,z)$ via the mass-to-light ratio $\Upsilon$,
$\rho_*(\varpi,z) = \Upsilon j_*(\varpi,z)$.  In this paper (as in
most previous studies), $\Upsilon$ will be considered a constant,
although in general, a spatially-dependent $\Upsilon$ could be used to
represent the contribution of a dark halo or a radially-varying
stellar mass-to-light ratio.

Obtaining $j_*$ from the observed surface brightness profile is an
under-determined problem for axisymmetric galaxies except when the
symmetry axis lies in the plane of the sky \cite{GerhardB96,
Rybicki87, RomanowskyK97}.  But galaxies in which the mass is
stratified on similar concentric ellipsoids do have unique
deprojections provided the inclination angle $i$ is known.  In general
we would obtain $\rho(\varpi,z)$ via a non-parametric deprojection of
the observed surface brightness profile (Merritt \& Tremblay 1994,
Merritt et al. 1997).
In what follows, the focus is on the indeterminacy resulting from
incomplete kinematical information and we will assume the freedom to
specify a unique functional form for $\rho_*(\varpi,z)$.

The gravitational potential is assumed to be of the form
\beq
\Phi(\varpi,z) = \Phi_*(\varpi,z) - G\mh/(\sqrt{\varpi^2+z^2}),
\label{eq:phi_tot}
\eeq
where $\Phi_*(\varpi,z)$ is the potential derived from the stellar
luminosity profile and the second term is the contribution from a
central black hole.  An efficient way to evaluate $\Phi_*(\varpi,z)$
is via a truncated multipole expansion \cite{vanAlbada77}: 

\begin{eqnarray}
\Phi_*(r,\theta) & = & -2\pi G \sum_{l=0}^{l_{\rm max}}{P_l(\cos\theta)} \times\\
& &\left[
{\frac{1}{r^{l+1}}}\int_0^r{ \rho_l(a)a^{l+2}{\rm d}a} +
r^l\int_r^\infty{\rho_l(a){\frac{{\rm d}a}{a^{l-1}}}}\right] \nonumber,
\label{eq:exp}
\end{eqnarray}

Expressions for the forces in cylindrical coordinates are easily
derived from equation (\ref{eq:exp}).  The density distribution is
required on a grid in $(r,\theta)$.  Since all real elliptical
galaxies have moderate to steep central density cusps, the radial grid
is chosen to be logarithmically spaced.  The potential between grid
points is evaluated by bicubic spline interpolation.

It proved convenient to choose an analytic form for the luminosity density.
Since the simulated data described
below were generated from a 2I model of M32 (see \S~3), we
adopted the parametrized form of the luminosity density used by 
\cite{vanderMarel98} (hereafter vdM98)
in the construction of this model:

\beq 
j_*(\varpi,z) = j_*(m) = j_0 (m/b)^\alpha [1 + (m/b)^2]^\beta
[1 + (m/c)^2]^\gamma, 
\label{eq:m32j}
\eeq

where $ m^2 = \varpi^2 + (z/q)^2$, and $\alpha = -1.435$, $\beta =
-0.423$, $\gamma = -1.298$, $b= 0.55$\arcsec, $c = 102.0$\arcsec, $q =
0.73 (i=90^\circ)$.

The potential due to the density distribution (\ref{eq:m32j}) can be
derived directly via Poisson's equation and the forces via numerical
quadrature.  We tested the accuracy of the multipole expansion scheme
by comparing the force evaluations with those obtained via quadrature.
We took $l_{\rm max}=6$ and set the inner radius of the grid at
$6\times 10^{-4}$\arcsec. 80 radial grid points and 8 polar grid
points were selected for the multipole expansion.  These tests showed
that the multipole expansion gives forces that have fractional errors
of $\sim 10^{-3}$ at the innermost radius, dropping rapidly with
increasing radius.  The use of the multipole expansion scheme results
in an approximately eightfold reduction in orbit integration times
compared with force evaluation via quadrature.

\subsection{The Orbit Library}
\label{sec:orblib}

All orbits in a steady-state axisymmetric Hamiltonian respect at least
two isolating integrals of the motion: the orbital energy $E$ and the
angular momentum $L_z$ about the symmetry axis. A non-resonant orbit
with only these two integrals would completely fill the region of the
meridional plane enclosed by the zero-velocity curve (ZVC). However
numerical studies e.g. \cite{Contopoulos60, Ollongren62, Richstone82}
show that most orbits also conserve a third integral, $I_3$, which
confines the orbit to a subset of the allowed meridional-plane
region. When the third integral is present, the orbit touches the ZVC
at a finite number of points.  Launching orbits from uniformly-spaced
points on the ZVC ensures a reasonable sampling of the third dimension
of phase space accessible to regular orbits.

Each orbit is integrated for a fixed number of periods and its
properties stored.  The number and nature of stored properties is
determined by the available data.  Since the purpose of generating the
orbit library is to determine the linear combination of orbits that
best reproduces the data, we need to ``observe'' each orbit under
conditions as close as possible to the conditions under which the data
were taken.  This involves convolving the intrinsic orbital properties
with the seeing function, as well as averaging over the observed slit
width and aperture size. The result is a set of quantities associated
with the orbits that can be linearly co-added and compared with the
data, without any need for interpolation.  In the remainder of this
section we describe the various steps in the generation of the orbit
library.

\subsubsection{Orbital Initial Conditions}
\label{subsubsec:orbinitial}

Our choice of orbital initial conditions is similar to that of vdM98
and Cretton et al. (1999).  We first select a radial grid of $N_E$ points
logarithmically spaced from $\varpi_{\rm min}$ to $\varpi_{\rm max}$;
for the mass model of equation (\ref{eq:m32j}), we took $\varpi_{\rm
min} = 5\times 10^{-4}$~\arcsec and $\varpi_{\rm max} = 7.5\times
10^{3}$~\arcsec.  At each radial grid point $\varpi_i$, the energy of
the circular orbit of radius $\varpi_i$ is $E_i =
(1/2\varpi_i)\partial \Phi/\partial \varpi_i + \Phi(\varpi_i,0)$, thus
defining the energy grid.  The maximum allowed angular momentum at
energy $E_i$, $L^i_{\rm max}$, is determined by the angular momentum
of a circular orbit. At each energy we choose $N_J$ regularly-spaced
values in $L_z$ on the open interval (0, $L^i_{\rm max}$)
(i.e. excluding $L^i_j=0$ and $L^i_j = L^i_{\rm max}$, which
correspond to radial and circular orbits respectively). This grid only
selects orbits with one sense of rotation about the symmetry axis, but
orbits with the opposite sense of rotation are trivially obtained by
flipping the sign of the velocity.  For each pair $(E_i,L^i_j)$ we
then compute the ZVC, the curve on the meridional plane where the
effective potential is zero:

$$
\Phi_{\rm eff} = \Phi(\varpi,z) + {\frac{1}{2}}{\frac{L_z^2}{\varpi^2}} = 0.
$$

The third quantity chosen to define an orbit is the angle $\beta$
between the major axis ($x$) and the line joining the origin and a
point on the ZVC.  We select $N_\beta$ equally-spaced angles $\beta$
in the open interval $(0, \pi)$.  In the tests described below, we
computed for each mass model a library with $(N_E, N_L,N_\beta) = (62,
9, 8)$ for a total of $\sim 4464$ orbits having one sign of rotation,
or $8928$ orbits overall.

Orbits were integrated in the meridional plane using an explicit
Runge-Kutta integrator of order 8(5,3) due to (Hairer \& Wanner 1993)
with adaptive step size control but which give dense output at equally
space time intervals. The integration interval ($N_{\rm period}$) was
taken to be 200 periods of the circular orbit at each energy and
orbits were sampled at $N_{\rm step} = 100$ equally-spaced time steps
during each orbital period.  Orbits were launched from the ZVC with
initial velocities $v_\varpi = v_z =0$. At the end of the integration
the energy of the orbit was always conserved to a (relative accuracy)
of better than $1\times 10^{-5}$. While integrations were carried out
in the meridional plane, we require the orbit in Cartesian coordinates
in order to compare with the observed data. Cartesian coordinates
$(x,y,z,v_x,v_y,v_z)$ were computed by assuming a random azimuthal
angle $0 \le \phi \le 2\pi$ at each time step and $v_\phi(t) =
L^i_j/\varpi(t)$. Unlike other authors e.g. \cite{Cretton99, Verolme02} we do not see the need to ``dither'' the orbits to create
packets of orbits. Also, unlike these authors we compute the forces
precisely (from the multipole expansion routine) at each point in the
orbit rather than interpolating from forces stored on a grid in
$(\varpi, z)$. Once the orbit is integrated in the potential the
observed properties of the orbit need to be transformed to the correct
viewing angle based on the assumed inclination $i$ of the model; this
gives an additional set of coordinates
$(x^\prime,y^\prime,z^\prime,v^\prime_x,v^\prime_y,v^\prime_z)$, with
$x^\prime$ and $z^\prime$ coinciding with the projected major and
minor axes respectively and $v^\prime_y$, the observed line-of-sight
velocity.

\subsubsection{The Storage Grids}

The orbital properties are stored on three kinds of grid, depending on
the type of observational constraint. These storage grids are similar
to those used by other authors \cite{Rix97, vanderMarel98, Cretton99, Verolme02}.

To reproduce the known mass distribution of the model
(self-consistency constraints), we store the orbital contribution to
the mass of each cell on a grid in the $(r,\theta)$-plane.  We use 20
logarithmically-spaced radial bins and 16 equally-spaced bins in
$\theta$ ($0 \leq \theta \leq \pi/2$).  For the M32 mass model
described above, the lower and upper radial grid points were at $\sim
5\times 10^{-4}$\arcsec and 102\arcsec. At each time step the orbital
position ($\varpi$, $|z|$) determines the cell to which a fractional
weight $\delta$ is added.  The total mass contributed by the
$\alpha$th orbit to the grid cell centered on $(r,\theta)$ is a sum of
all the fractional weights and is represented by
$m^{\alpha}_{r\theta}$.

The orbital kinematics are stored on 3-D grids in the projected
coordinates $(x^\prime, z^\prime, v^\prime_y)$. Each set of
observations (defined by different seeing, aperture locations etc.)
requires a separate grid.  The grids themselves are square in the
$x^\prime-z^\prime$-plane with outer boundaries set by the outermost
observed aperture. For the models in this paper the typical grid
consisted of $267\times 267$ pixels  with the bin width equal
to $\sim 1/8$ the FWHM of the PSF (or seeing in the case of ground
based data). So for instance for all data from the HST (FOS and STIS)
the orbit libraries were stored on grids with pixel width
0.0125\arcsec\ whereas for ground based CFHT data (e.g. Bender et al.~1996)
the pixel width was 0.038\arcsec.  The grid in the velocity dimension
has 107 points in the range [-800\kms, 800\kms~] or a velocity
sampling of 15.1\kms. This is smaller than the velocity scale of
the STIS spectrograph ($\sim$ 19\kms\, per pixel at $\sim 8500$~\AA\, 
or a wavelength scale of 0.56~\AA\, per pixel). In general it was found
necessary to use a velocity range which is at least $\pm 4-6 \times
\sigma_{\rm max}$, where $\sigma_{\rm max}$ is the largest observed
velocity dispersion.

It is standard practice to generate orbit libraries for a single value
of the mass-to-light ratio $\Upsilon_0$ and to generate libraries for
all other $\Upsilon$ values by scaling the velocities by a factor
$\sqrt{\Upsilon/\Upsilon_0}$ (e.g. vdM98; Cretton et al. 1999).  We will
refer to the library generated using $M/L=\Upsilon_0$ as the ``primary
library'' for each value of $\mh$.  It is important that the choice of
$\Upsilon_0$ be determined by a prior estimate of the best-fit value
of $M/L$ (based on e.g. 2I or spherical models). If the velocities in
the primary library are stored on a grid with range $[-v_0, v_0]$ and
grid spacing $\delta v_0$, the scaled velocities for any other
$\Upsilon$ will have a range $[-\sqrt{\Upsilon/\Upsilon_0}v_0,
\sqrt{\Upsilon/\Upsilon_0}v_0]$ and velocity spacing of
$\sqrt{\Upsilon/\Upsilon_0}\delta v_0$. The value of $v_0$ must be set
by the smallest $\Upsilon_{\rm min}$ for which the model will
subsequently be scaled: $\sqrt{\Upsilon_{\rm min}/\Upsilon_0}v_0
\simeq 4\sigma_{\rm max}$, and the value of $\delta v_0$ should be set
by the largest $\Upsilon_{\rm max}$ to which the model will be scaled:
$\sqrt{\Upsilon_{\rm max}/\Upsilon_0}\delta v_0 \simeq \Delta v_{\rm
obs}$ where $\Delta v_{\rm obs}$ is the velocity sampling of the
highest-resolution spectrographic data set.  Carelessness in this
regard can lead to spuriously poor fits to data at low and/or high
values of $\Upsilon$.

Since we store the orbit at equal time intervals, each time the
$\alpha$th orbit passes through a cell centered on ($x^\prime,
z^\prime, v_y^\prime$) it adds a  constant fractional weight $\delta = 1/(N_{\rm
period}\times N_{\rm step})$ in that cell. At the end of the
integration we store the total weight $\omega^\alpha_{x^\prime
z^\prime v_y^\prime}$ contributed by this orbit to each cell. In
practice it was found to be better to construct the orbital LOSVDs
using a kernel density estimator (with a kernel width of 2.5$\times
\delta v_0 \sim 38$\kms~) rather than by simple binning in $
v_y^\prime$ since this results in smoother LOSVDs without compromising
velocity resolution of the orbital LOSVDs.  This practice
significantly improves the accuracy and speed of fitting the observed
LOSVDs.

A final grid in the $(r,\theta)$- plane is used to store 3-D
kinematical properties of the orbits. We store the density weighted
(un-centered) first and second moments of the LOSVDs in spherical
polar coordinates: $\overline{\rho v_r}$, $\overline{\rho v_\phi}$,
$\overline{\rho v_\theta}$ and $\overline{\rho v_r^2}$,
$\overline{\rho v_\phi^2}$, $\overline{\rho v_\theta^2}$. These 6
quantities as well as $\rho$ the average density (in the cell) are
computed and stored in each of the 20 radial and 16 polar cells
described above. These quantities are not used in fitting the data but
are useful for analyzing the properties of the resulting models.

\subsubsection{PSF-Convolution}

Convolution with the point spread function (PSF) is essential when
comparing the orbit libraries
with the observations.  The choice of Cartesian grids in
$(x^\prime, z^\prime, v^\prime_y)$ for storing the kinematical data is
driven by the fact that PSF convolution is most easily carried out in
Fourier space via standard Fast Fourier transforms (FFTs).

For this paper we assume that all PSFs are circularly symmetric
Gaussian (or multiple Gaussian) with FWHM given by the observed
seeing.  Bower et al. (2001) have shown that the STIS/CCD PSF
at $\sim 8500$~\AA\ has a FWHM = 0.079\arcsec\ with a broad asymmetric
wing on one side. This ring represents the first Airy ring in the PSF
and probably arises from misaligned optical elements. Bower et al.
also carried out tests with synthetic spectra to show that a symmetric
model PSF obtained by folding and averaging the true PSF about the
center reproduces the observed data to within the errors. They found
that even when noise was not added to the spectrum, the kinematic
measurements from the model PSF and the observed PSF were not
statistically different.  We therefore use a PSF which is a circular
Gaussian with FWHM of 0.1\arcsec\ for both the PSF convolution with
the orbit library, as well as for generation of the simulated-data.

PSF convolution with a seeing function correlates the data in the two
spatial directions but does not affect data in the velocity
direction. Therefore PSF convolution is carried out separately for
each 2-D velocity slice of each of the ($x^\prime, z^\prime, v_y^\prime$)
grids. PSF convolution redistributes the orbital weights and we now
represent the weight due to the $\alpha$th orbit in the bin centered
on ($x^\prime, z^\prime, v_y^\prime$) by
$\tilde{\omega}^\alpha_{x^\prime z^\prime v_y^\prime}$.

In order to properly scale the orbital LOSVDs observed though
different apertures, it is essential to know the total flux observed
through each aperture. In general this information is not available
from the kinematical data.  We therefore compute this from the model
density distribution on a Cartesian grid with the same spatial
resolution as each of the kinematic storage grids. These projected
mass grids are also convolved with the appropriate PSFs. The resultant
projected mass in each grid cell is represented by $\tilde{s}^{\rm
obs}_{x^\prime z^\prime}$.

(PSF convolution was carried out using a FFT routine originally
written by Norman and Brenner of MIT Lincoln Labs in 1968 and modified
for the current problem and kindly made available by R. van der Marel.)

\subsubsection{``Observing'' the Orbit Library }
\label{sec:observables}

After each velocity slice of the Cartesian storage grid and the
Cartesian projected mass grid is convolved with the PSF, the kinematic
properties of each orbit (and its projected mass) are ``observed''
through the same set of apertures as the data. Following Rix et
al. (1997) and Cretton et al. (1999) we use a simple scheme to compute
the contribution of each pixel of a storage bin to each aperture. Each
pixel contributes a fraction $\tau_{x^\prime z^\prime l}$ to the $l$th
aperture, where $0 \leq \tau_{x^\prime z^\prime l} \leq 1$ depending
on whether the pixel centered on $(x^\prime, z^\prime)$ lies entirely
outside the aperture, on the edge of the aperture, or entirely inside
the aperture. Since the positions and orientations of the apertures
relative to the Cartesian grids is fixed for all the individual orbits
these $\tau_{x^\prime z^\prime l}$ are computed at the start of the
orbit library program and stored. The un-normalized LOSVD of the
$\alpha$th orbit as seen through the $l$th aperture is then obtained
simply by 
\beq 
N^\alpha_l(v_y^\prime) =
\sum_{x^\prime,z^\prime}{\tilde{\omega}^\alpha_{x^\prime z^\prime
v_y^\prime} \cdot \tau_{x^\prime z^\prime l}}.
\label{eq:losvd}
\eeq 
The total orbital mass
contribution to the $l$th aperture is 
\beq 
m^\alpha_l =
\sum_{x^\prime,z^\prime,v_y^\prime }{\tilde{\omega}^\alpha_{x^\prime
z^\prime v_y^\prime} \cdot \tau_{x^\prime z^\prime l}}. 
\eeq
\noindent
Finally, as noted earlier, the observed flux through each aperture is
information that is not generally available from the data but is
required for proper scaling of the LOSVDs. We therefore compute the
``observed'' mass in each aperture $M^{\rm obs}_l$ from the
theoretical surface density profile of the model via
\beq 
M^{\rm obs}_l =
\sum_{x^\prime,z^\prime}  {\tilde{s}^{\rm obs}_{x^\prime
z^\prime} \cdot \tau_{x^\prime z^\prime l}}. 
\eeq

\subsection{Constructing the Model}
\label{sec:lpp}

The construction of a 3I model to fit the constraints now consists of
finding a weighted superposition of the orbits that best reproduces
both the assumed model stellar density distribution $\rho(\varpi,z)$
and the observed LOSVDs, or some representation of the LOSVD. If there
are $N_c$ is total number of observational constraints (mass and
velocity), and $N_o$ is the number of orbits, we minimize the mean
square deviation in the quantity $\chi^2$, where 
\beq 
\chi^2 = {\frac{1}{N_{\rm c}}}\sum_{m=1}^{m=N_{\rm c}}\left(D_m -
\sum_{\alpha=1}^{\alpha = N_{\rm o}} \gamma^\alpha B^{\alpha}_m
\right)^2,
\label{Eq:generic_chisq}
\eeq
subject to a basic set of non-negativity constraints:
\beq
\gamma^\alpha >0.
\label{eq:nonnegtive}
\eeq

In the set of equations above $\gamma^{\alpha}$ is the weight assigned
to the $\alpha$th orbit, $D_m$ are the $N_c$ observational constraints
and $B^{\alpha}_m $ is the contribution of the $\alpha$th orbit to the
$m$th constraint. The matrix elements $D_m$ and $B^{\alpha}_m$ are
replaced by the various observable quantities described in
\S~\ref{sec:observables} as well as other quantities that are required
to construct the self-consistent model, such as the mass $M^{\rm
obs}_{r\theta}$ in each cell. This is not an observed quantity but is
derived from $\rho(\varpi,z)$.  The corresponding orbital masses
$m^\alpha_{r\theta}$ that are superposed are weighted by
$\gamma^{\alpha}$ such that, 
\beq 
M^{\rm obs}_{r\theta} =
\sum_{\alpha}\gamma^{\alpha}m^{\alpha}_{r\theta},
\label{eq:mass_constraints}
\eeq 

In principle one can attempt to fit all the observed data as well as
the mass (self-consistency) constraints to within numerical
precision. In practice, the observed LOSVDs (and quantities derived
thereof) have finite errors and there is nothing to be gained by
attempting such precision in the model fits.  Following standard
procedure, we account for the errors in different quantities by
dividing both the observed data and the corresponding quantity in the
orbit library by the error on the observed data. 

Since the self-consistency (mass constraints) in
eq.(\ref{eq:mass_constraints}) are derived and not observed
quantities, there are no ``observed errors'' on them. It is therefore
possible to arbitrarily set the relative weighting of the kinematic
constraints and mass constraints (which have essentially infinite
accuracy). Instead of an error we use a constant scaling factor
($1/\delta M$) which sets the weight of the mass constraints relative
to the kinematical constraints. For each data set one needs to
experiment to determine the scaling factor that gives a consistently
good fit to the mass constraints for all input parameters while
satisfying the kinematic constraints.  (Note that unlike Rix et
al. (1997) we do not explicitly include aperture mass constraints in
the objective function because here too the errors in the aperture
masses are unknown. If we were to include them, this would introduce
yet another free scaling factor. Also, unlike Rix et al.  we do not
separately fit the surface density distribution, since this is
automatically guaranteed by an accurate fit to the mass distribution.
We have found that it is generally possible to fit the meridional
plane masses to a fractional accuracy of $\sim 10^{-2} - 10^{-5}$ over
the entire $\mh-\Upsilon$ plane and this always guarantees a fit to
the projected mass (or equivalently surface brightness distribution)
with error of less than 1\%.)

The second set of constraints to be fitted are the kinematic
constraints, consisting of the LOSVDs observed through each
aperture. The un-normalized orbital LOSVDs given in equation
(\ref{eq:losvd}) can be linearly superposed to obtain a fit to the
observed LOSVDs:

\beq
N^{\rm obs}_l(v_y^\prime) = \sum_{\alpha}\gamma^{\alpha}N^\alpha_l(v_y^\prime) 
\eeq

Since LOSVDs are often approximately Gaussian in shape, it is common
practice to represent the observed LOSVDs through a truncated
Gauss-Hermite series. The highest quality spectra can yield useful
Gauss-Hermite moments up to order 6; fitting of moments up to order 4
is now standard. We follow the method suggested by \cite{Rix97} to
linearly superpose {\it mass-weighted} orbital GH moments that are
linear in the orbital LOSVDs and refer the reader to this source for
details. The observed kinematic data do not include information on the
lowest order moment of the LOSVDs ($h_0$ or $\gamma_0$), or the total
flux through each aperture ($M^{\rm obs}_l$)

Previous authors have fitted either the GH moments (e.g. \cite{Rix97};
vdM98; \cite{Cretton99}; \cite{CrettonB99}; \cite{Cappellari02}; \\
 \cite{Verolme02}) or the entire LOSVD \cite{Gebhardt00a, Bower01}.  In
principle it is possible to fit a combination of both kinds of
constraint. It is generally observed that LOSVDs are likely to deviate
strongly from a Gaussian (due to high-velocity wings) only in a few
apertures close to the center.  For these apertures it may be
important to fit the full LOSVD.  If the LOSVDs are explicitly fitted
in the central apertures labeled by $l$, $1\le l\le l_1$, and the
lowest few GH moments are fitted elsewhere, $l_1+1\le l\le l_{\rm
max}$, then the problem of fitting the data via a linear superposition
of the orbits can be viewed as a problem of minimizing an objective
function of the form

\begin{eqnarray}
\chi^2 N_c & = & {\sum_{r\theta}{\left[{{M^{\rm
obs}_{r\theta}-\sum_{\alpha}{\gamma^{\alpha}m^{\alpha}_{r\theta}}}}
\over{\delta M}\right]^2}} \nonumber\\
& & +\sum_{l=1}^{l_1}{\left[{{N^{\rm obs}_l(v_y^\prime)-\sum_{\alpha}{\gamma^{\alpha}N^{\alpha}_l(v_y^\prime)}}\over{\errNol}}\right]^2}\nonumber\\
& & + \sum_{l = l_1+1}^{l_{\rm max}}\sum_{i=1}^{h_{\rm max}}{\left[{{M^{\rm obs}_l h^{\rm obs}_{li}-\sum_{\alpha}{\gamma^{\alpha}{\cal
H}^{\alpha}_{li}}}\over{\Delta(M^{\rm obs}_l h^{\rm obs}_{li})}} \right]^2}.
\label{Eq:Chisq}
\end{eqnarray}

The mass-weighted Gauss Hermite moments ${\cal H}$ are given by
\begin{eqnarray}
{\cal H}_{li}^{\alpha}&= &2\sqrt{\pi}\int_{-\infty}^{\infty}  
N_l^{\alpha}(v_y^\prime)g(w)H_i(w)dv,  \\ 
i & = & 1,h_{\rm max}, \nonumber\\
g(y) &= & (2\pi)^{-1/2} e^{-y^2/2},\nonumber\\
w &=& (v-V_l)/\sigma_l.\nonumber
\end{eqnarray}
Typical values of $h_{\rm max}$ are 4 or 6.  We are free to multiply
each pair of terms inside the same square brackets in the objective
function by a constant factor, e.g. a scaling factor or an inverse
error. In equation \ref{Eq:Chisq} we have multiplied each term by an
inverse error for illustration. This gives equal weight to each of the
different terms in equation~\ref{Eq:Chisq}.

Minimization of the objective function was carried out using two
different software packages: the quadratic programming algorithm
E04NCF of the NAG libraries, and a non-negative least squares (NNLS)
routine  \cite{Lawson95}.  The two algorithms
gave similar results; for all models described below we present the
fits obtained using the NAG routine.

Unless otherwise noted, we use the symbol $\chi^2 $ to represent the
objective function including {\it all quantities included in the fit}
and not just e.g. the kinematical constraints.  Since the objective
function includes errors in the measured quantities, $\chi^2$ as we
define it should be loosely interpreted as a reduced $\chi^2$,
although as we discuss below, that interpretation is problematic.

\subsection{Regularization}
\label{sec:regularize}

One disadvantage of an orbit-based approach to model building is that
the solutions are extremely unsmooth.  One source of this lack of
smoothness is the discrete way in which phase space is sampled.  But
even more important is the inherent ill-conditioning of the
self-consistency problem \cite{Merritt93b}.  A single orbit, which
represents a delta-function in integral space, covers a finite region
in configuration space.  Deriving the integral-space density from the
configuration space density is therefore a deconvolution problem, and
deconvolution has the property of amplifying errors or incompleteness
in the data.  Even a highly noisy set of orbital weights can generate
a smooth configuration-space density, and there are many more noisy
solutions than smooth ones.  This effect actually becomes worse as the
number of orbits is increased since a fine grid is better able than a
coarse grid to represent high-frequency fluctuations \cite{Phillips62}.

Lack of smoothness is an inconvenience when plotting deprojected
quantities, and for this reason it has become standard practice to
couple Schwarzschild's technique with some sort of ``regularization''
scheme to enforce smoothness (e.g. Richstone \& Tremaine 1988; Cretton
et al. 1999; Gebhardt et al. 2003). But a deeper worry is that the
ill-conditioning might lead the optimization algorithm toward a 
non-smooth solution that has no
smooth counterpart.  If imposing smoothness on a numerical solution
causes it to depart strongly from self-consistency, one would conclude
that no solution continuous in the phase-space variables exists, and
that the apparent self-consistency is a numerical artifact associated
with the discretization.  Merritt \& Fridman (1996) first investigated
this question in the context of Schwarzschild modeling of triaxial
galaxies; they found that their nonsmooth solutions had the same,
average properties as solutions for which smoothness was imposed.  On
the other hand, Jalali \& de Zeeuw (2002) found in modeling
scale-free disks that spurious solutions could be generated by using a
number of orbits that was large compared to the number of mass
constraints.

In the context of potential estimation, we need to check that the
indeterminacy in quantities like $\mh$ is not an artifact of noise in
the solutions.  For instance, it is possible that solutions with the
``wrong'' $\mh$ are much noisier than solutions with the ``true''
$\mh$, or that the range of indeterminacy is strongly dependent on the
level of smoothing.

A standard way to regularize is by adding a penalty term
to the objective function (\ref{Eq:generic_chisq}):
\beq 
\chi^{\prime\, 2} = {\frac{1}{N_{\rm
c}}}\sum_{m=1}^{m=N_{\rm c}}\left(D_m - \sum_{\alpha=1}^{\alpha = N_{\rm
o}} {\gamma^\alpha B^{\alpha}_m} \right)^2 + \lambda \sum_{\alpha =
1}^{\alpha = N_{\rm o}} {P(\gamma^\alpha)}
\label{Eq:regularize_chisq}
\eeq
where $P(\gamma^\alpha)$ is defined to be large and positive 
when the solution is unsmooth \cite{Phillips62,Thikonov63}. 
A number of choices are possible for $P(\gamma^\alpha)$,
depending on the definition of ``smoothness.''
Here we follow Merritt \& Fridman (1996) by adopting
``zeroth-order'' regularization:
\beq
P(\gamma^\alpha) = (\gamma^\alpha)^2
\eeq
(e.g. Miller 1974) which has the effect of filtering
fluctuations on scales shorter than some maximum value
determined by the smoothing parameter $\lambda$.
Models with $\lambda = 0$ have no regularization 
and models with $\lambda \rightarrow \infty$
are characterized by uniform orbital weights.

Having obtained a solution by minimization of equation
(\ref{Eq:regularize_chisq}), one would like to measure the degree of
smoothness.  The simplest way would be via $P(\gamma^\alpha)$, with
$\gamma^\alpha$ the orbital weights corresponding to the smoothed
solution.

Alternatively one can attempt to measure the degree of smoothness in
phase space of the function $\gamma(E,L_z,\beta)$ the orbital weights
on the grid of orbital initial conditions described in
\S~{\ref{subsubsec:orbinitial}}.  Following Cretton et al. (1999) we
compute the second-divided difference (in place of second derivative)
of the dimensionless function
$\gamma(E,L_z,\beta)/\gamma_0(E)$. $\gamma_0(E)$ the ``reference
weights'' and are a rough approximation to the energy dependence of the
model. Following Rix et al. (1997) we employ the simplest possible
regularization by setting all the $\gamma_0(E) = 1$ and characterize
the smoothness via the noise parameter:
\begin{eqnarray}
 \Pi  & =& {\frac{1}{N_{R}}}\sum_{i= 1}^{N_R}\\
      &  & {\left( {\frac{\partial^2
\gamma(E,L_z,\beta)}{\partial E^2}} + {\frac{\partial^2
\gamma(E,L_z,\beta)}{\partial L_z^2}} + {\frac{\partial^2
\gamma(E,L_z,\beta)}{\partial \beta^2}} \right)_i \nonumber} 
\label{eq:noisepi}
\end{eqnarray}
where $\partial^2 \gamma(E,L_z,\beta)/\partial E^2$ etc. 
represent the second divided differences of the weights of
adjacent orbits in the space $(E,L_z,\beta)$, and $N_{R}$ is the
number of interior grid points for which a second divided difference
can be computed (e.g. Cretton et al. 1999 ).  

We have used both the quantities $P(\gamma^\alpha)$ and $\Pi$ to
quantify the degree of noise and find little difference in the
results. Since the quantity $\Pi$ has been used in other studies and
is more physically meaningful we use it to represent the degree of
smoothness of our models in the discussion in \S~7.

It is interesting to note that for any smoothed model the
contributions from different parts of phase space to the total noise
($\Pi$ in eq. \ref{eq:noisepi}) depend primarily on energy $E$
remaining roughly constant at all values of $(L_z, \beta)$ at a given
energy. The noise in phase space is smallest at small energies and
increases slowly with radius (energy) reaching a maximum at $\sim$ the
35th energy level dropping slowly thereafter.

\bigskip
\section{Indeterminacy of the Three-Integral Problem}
\label{sec:indeterm}

Before discussing the results obtained by applying our 3I modeling
algorithm to simulated data, we review the reasons why we expect the
potential estimation problem to be under-determined in the axisymmetric
geometry, given the sorts of data (kinematical quantities measured
along multiple long slits) that we are dealing with here.

Consider first the spherically symmetric case.  Deprojection of
$\Sigma(R)$ yields $j(r)$, the luminosity density, uniquely; given
values for ($\Upsilon$, $\mh$), the mass density $\rho(r)$ and
potential $\Phi(r)$ are also known.  Suppose that the stellar
distribution function $f$ is isotropic, $f=f(E)$.  Then Eddington's
formula gives the unique $f$ that reproduces $j(r)$ in this $\Phi(r)$,
and corresponding to this unique $f$ is a particular RMS velocity
profile $\sigma^2(r)=\int f(E) v^2 d{\bf v}$.  Changing $\Phi$ will
change both $f$ and $\sigma$ in well-defined ways, so that the
goodness-of-fit of $\sigma(r)$ to the observed RMS velocities will
vary continuously with the parameters $(\Upsilon,\mh)$ that define the
potential.  Therefore, there will generally exist a best-fit (minimum
$\chi^2$) set of parameters for any kinematical data set.  This has
been illustrated in numerous studies  \cite{The86, Little87, Kulessa92, MerrittT93}. 

Suppose next that the stellar distribution function has the more
general form $f=f(E,L^2)$ with ${\bf L}$ the angular momentum per unit
mass.  There are now many functions $f(E,L^2)$ that can reproduce a
given $j(r)$ in a specified $\Phi(r)$, since $j(r)$ is a projection
over velocities of $f(E,L^2)$ and different 2D $f$'s can have exactly
the same 1D projection.  The same is true if additional moments of the
distribution function (e.g. $\sigma(r)$) are added as constraints:
many 2D functions $f$ are still able to reproduce a finite set of such
1D constraint functions.  This means that one has the freedom to vary
$f$ along with $\Phi$ in order to maintain the goodness-of-fit to the
data, subject only to the constraint that $f$ be non-negative.  The
result is an indeterminacy in the parameters that define the
potential: in general, there will be a range of potentials for which
$f$ can be adjusted such that the fit to the data is equally good, and
no ``best-fit'' potential can be found.  The indeterminacy of
potential estimation in the spherical geometry has been extensively
demonstrated (e.g. \cite{DejongheM92}; \cite{Merritt93a},b;
\cite{MerrittS93}).  These studies document that the range of allowed
potentials -- i.e.  potentials consistent with a non-negative
$f(E,L^2)$ given a finite set of data constraints like $\Sigma(R)$ and
$\sigma^{obs}(R)$ -- can be extremely wide.

Consider next the axisymmetric case.  Inversion of $\Sigma(X,Y)$ can
give $j(r,\theta)$ uniquely if the galaxy is known to be edge-on;
otherwise there is an indeterminacy in $j$ \cite{Rybicki87,
GerhardB96}.  We ignore that indeterminacy here and assume that
$j(r,\theta)$ is precisely known.  Suppose first that $f$ is
restricted to its simplest possible form consistent with axisymmetry,
$f=f(E,L_z)$.  Just as in the spherical isotropic case, there is a
unique, 2I $f$ that reproduces a given $j(r,\theta)$ in a specified
$\Phi(r,\theta)$ \cite{LyndenBell62, Hunter75, Dejonghe86}.
Furthermore this unique $f$ is associated with unique values for the
mean square velocity at every point in the projected image.  Varying
$\Phi$ will force both $f$ and its associated velocity field to vary,
hence once expects to find a single set of values ($\Upsilon,\mh)$
that provide the best fit to the measured velocities.  This has been
verified in a number of 2I modeling studies (e.g. Binney et al. 1990;
\cite{Dehnen95, Qianetal95, Magorrian98}).

In the general axisymmetric case, $f$ is a function of three
variables, $f=f(E,L_z,I_3)$ (assuming as above that all orbits are
characterized by three isolating integrals).  Just as in the
anisotropic spherical case, there are now many functions
$f(E,L_z,I_3)$ that can reproduce a known $j(r,\theta)$ in a specified
$\Phi(r,\theta)$, since many 3D functions $f$ project to the same 2D
function $j$.  The same will be true if to $j$ are added a finite set
of 2D data constraints, such as the mean square velocity measured over
the image of the galaxy.  The argument that was made above in the
anisotropic spherical case then applies to the 3I axisymmetric case:
changes in the assumed form of $\Phi(r,\theta)$ can generally be
compensated for by changes in $f$ so as to leave the fit to any finite
set of 2D data constraints precisely unchanged, and one expects to
find a range of potentials over which the goodness-of-fit to the data
is constant.  The extent of this constant-$\chi^2$ region is
determined by the requirement that $f\ge0$; if the potential is made
sufficiently extreme, the only $f$'s that can reproduce the data will
be negative somewhere in phase space, and the fits of non-negative
$f$'s to the data will begin to suffer.

In the anisotropic spherical case, it is generally believed that
measuring the LOSVDs at a large enough set of radial positions can
uniquely constrain the potential.  Numerical experiments seem to bear
this out \cite{Merritt93a, Gerhard93} although only a small set of
cases have been tested and no general theorems have been proved.
Similarly in the 3I axisymmetric case, it is hoped 
(e.g. \cite{Cappellari03}) that sufficiently good, 2D data will
uniquely constrain both $\Phi(r,\theta)$ and $f(E,L_z,I_3)$.  This is
at the present time only a hypothesis, and given the non-linear
relation between the data and the potential, we expect that a given
data set will either under-, or over-constrain the potential; a
precise match between the information content of the data and
potential seems difficult to achieve.

We stress that the indeterminacy discussed here is mathematical, not
statistical, in nature, and is not due simply to the fact that
operations like deprojection are ill-defined when data are noisy or
incomplete (although those factors may contribute to the indeterminacy
e.g. \cite{CrettonE03}).  This means that any statistic like $\chi^2$
that measures the mean deviation between the data and the model will
generally be precisely constant over finite regions of parameter space
-- regions in which the data functions predicted by the model are
unchanged as the model parameters are varied.  We suggest that a
sensitive test of the quality of a 3I modeling algorithm is its
ability to reproduce such perfectly-flat $\chi^2$ plateaus, since any
limitations in the flexibility of the algorithm will keep it from
reproducing some $f$'s as well as others, resulting in spurious minima
in $\chi^2$.  For instance, if a 3I algorithm were written in such a
way that it could only reproduce the subset of $f$'s satisfying
$f=f(E,L_z)$, one would always find a unique minimum in
$\chi^2(\Upsilon,\mh)$.

\section{A Test Case: A 2I Model of M32}
\label{sec:2I_model}

We would like to test our algorithm against a reasonably realistic,
axisymmetric galaxy model whose properties are precisely known.
For this purpose we constructed an axisymmetric two-integral
(2I) model, $f=f(E,L_z)$, with properties very similar to those
of models that have been fitted in the past to data from M32.
In this section we describe the construction of that model and
the way in which we generated simulated ``data sets'' from it.

We constructed 2I models using the Hunter \& Qian (1993) (HQ)
prescription to derive the even part of the distribution function from
a given mass model.  The mass model was represented by a sum of 3D
Gaussian functions using the Multi Gaussian Expansion (MGE) method
(Monnet et al. 1992; Emsellem et al. 1994).  This method allows one to
obtain a simple analytic form for the potential; the HQ derivation is
also simplified due to the fact that the exponential form (Gaussians)
separates well in the complex plane.  Thus an analytical continuation
of the potential known only on the real axis is straightforward. (It
is important to note that while the MGE method described below is used
to generate the density profile for the pseudo-data and the orbit
library is constructed using the analytic density profile in
eq.~\ref{eq:m32j}, both density profiles agree extremely well with
each other.)

The 2I models were designed to give a good fit to all space-based and
ground-based observations of M32 available up to the year 2001.  These
data include long slit spectra along four position angles, and one
slit offset from the major axis, obtained with the WHT
\cite{vanderMarel94a}; CFHT spectra (Bender et al.~1996); HST/FOS
spectra at eight apertures close to the major axis (van der Marel et
al. 1997); and the HST/STIS spectra of Joseph et al. (2001).

A fit to the surface brightness distribution was obtained by applying
the MGE method to both a wide field and a high resolution I-band
image.  The wide field image, kindly provided by R. Michard and taken
at INT/PFCU, contained $382\times 575$ pixels (0.549 \arcsec/pixel);
the resolution was modest, $\gtrsim 2$ \arcsec FWHM.  The MGE fit was
first done directly on the wide field image to constrain the large
scale luminosity distribution, after masking any point sources
(e.g. stars).  The fit was found to be good down to 19.5
mag~arcsec$^{-2}$ with the sky becoming a problem at fainter levels.
The broadest Gaussian had a $\sigma$ of about 45\arcsec: this means
that at a radius of 100\arcsec, the luminosity of the model drops very
rapidly (exponentially).  Previous tests have shown that this should
not influence the central kinematics (Emsellem et al. 1999).  The
low-frequency components (Gaussians with $\sigma$s larger than
8\arcsec) of the original fit were then removed from the high
resolution image (in the case of M32 the WFPC2/F814W image was used
after proper normalization).  A fit was then performed against the
residuals using a 4-Gaussian approximation for the WFPC2 PSF in the
F814W filter. The resultant fit provides the deconvolved model for the
surface brightness at the very center (for more details see Emsellem
et al. 1999).  The final model for M32 consisted of 11, 2D Gaussian
components.  Since even the HST WFPC observations have a finite
spatial resolution which causes a spurious turnover in the central
density, the central luminosity profile was replaced by a power-law
component, or cusp.  This cusp was prescribed as in Emsellem et
al. (1999), with a power law slope of 1.5 ($j_*(r) \propto r^{-1.5}$)
and a Gaussian width of $0.05$\arcsec.

 The total energy of the model was kept constant when the cusp was
added and this additional component did not change the fit of the
surface brightness distribution in the central parts.  The even part
of $f$, $f_+\equiv {1\over 2}[f(E,L_z)+f(E,-L_z)]$, was then derived
for an assumed angle of inclination $i$, mass-to-light ratio
$\Upsilon$ and black hole mass $\mh$.  The simulated data sets
described below were derived from a model with $i=90^\circ$ (edge-on),
$\Upsilon_V=2$ and $\mh=2.625\times 10^6\msun$.  The odd part of $f$
was chosen following the prescription of Emsellem et al. (1999), by
flipping the direction of orbits with respect to the symmetry axis
until the best fit was obtained to the observed kinematics.  The
projected LOSVDs were then computed on a very fine grid (1600
logarithmically spaced points within the one quadrant of the central
15\arcsec).  Finally, the LOSVDs were convolved to take into account
the seeing and the instrumental PSFs and averaged over the apertures
(pixel sizes) appropriate to each set of simulated observations.  We
assume a distance to M32 of $0.7$ Mpc, as in earlier studies
(e.g. vdM98).

Two simulated data sets were constructed from this 2I model:\\ {\it
Data set A} was designed to simulate kinematical data obtained by STIS
on HST.  The 2I model was ``observed'' at STIS resolution (0.1\arcsec)
in 0.05\arcsec$\times$0.1\arcsec\ apertures from -1.5\arcsec\ to
1.5\arcsec\ along the major axis and the HST PSF was applied.  The
LOSVDs were extracted in each aperture and sampled at 5~\kms\
intervals.  These LOSVDs were then used to compute the projected
velocity $V$ and velocity dispersion $\sigma$ as well as the first six
GH moments at each aperture position.  In addition, the LOSVDs were
resampled at two other velocity spacings: 40\kms (comparable to that
velocity resolution of the STIS spectrograph $\sim~38$~\kms~) and at
100~\kms, corresponding approximately to the velocity resolution of
the FOS spectrograph (used to observe M32 by van der Marel et
al. [1997] and to observe NGC~3379 by Gebhardt et al. 2000a).\\ {\it
Data set B} was obtained by ``observing'' the 2I model with the same
set of apertures and PSFs as in the data compiled by vdM98 and used by
those authors in the construction of 3I models for M32.  These data,
consisting of combined data sets from the WHT, CFHT and FOS, are the
same data used in constructing our 2I model.
\setcounter{figure}{0}
\begin{figure}
\figurenum{1}
\label{fig:pseudo-setA}
\epsscale{1.}
\plotone{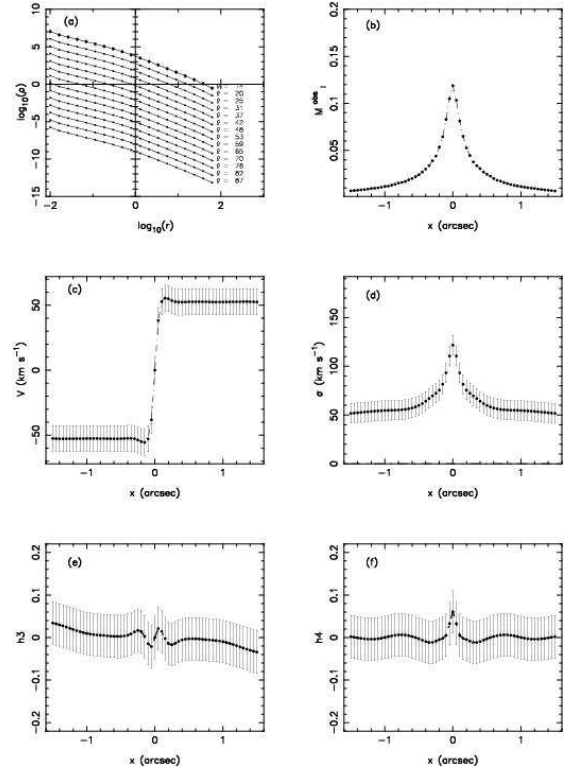}
\caption{
All mass and kinematical constraints for simulated data set A.
(a) The model density in a total of 266 cells at 16 radial intervals
and 14 polar angles ($\theta$ in degrees). The density is plotted
in arbitrary units (density profile for each polar angle is offset
from the previous angle by 1 unit). Error bars used in the actual fits
are plotted for $\theta = 14$ but are multiplied by a factor of 10
for visibility. (b) The projected (theoretical) mass in apertures
which is used to scale the GH moments; (c)-(f) $v_l, \sigma_l, h_3,
h_4$ with error bars used in the model fits.
}
\end{figure}

Since these are simulated data, there are no errors and no scatter in
 the data points. There are two ways in which ``pseudo-errors'' may be
 assigned to data points. First all velocities and velocity
 dispersions, and GH moments can be assumed to have a fixed error
 (e.g. we choose an error of $10$ \kms\ in $V$ and $\sigma$, and $h_3$
 and $h_4$ were assumed to have errors of 0.1).  Such error values are
 fairly typical of those associated with real HST/STIS data and CFHT
 data but somewhat larger than the errors associated with the WHT
 data. Alternatively the pseudo-data can be assumed to have the same
 errors at each point as the real data.
 
 In addition to error, real data have scatter. In the interest of
 keeping the number of free parameters to a minimum the pseudo data
 sets A and B do not have any scatter. This could affect the solution
 space by allowing models that are systematically different but not too
 far off to give equally good fits to the data, where one might have
 been harder to fit had there been appreciable scatter. 
 
 In order to introduce scatter into the pseudo data in a meaningful
 way we would need to run models for a variety of different levels of
 scatter to determine how scatter affects the results. Such a study is
 beyond the scope of this paper. However in order to ensure that the
 results are not purely an artifact of the ``pseudo'' nature of the
 data, in addition to these simulated data sets, we also applied our
 modeling algorithm to the actual kinematical data in vdM98.  We refer
 to these data as {\it data set C}.  Of course, data set C can not
 serve as a test of our algorithm since we do not know the true
 ``model parameters'' of M32!  However these data do allow us to
 compare our results with those of vdM98, and to test the sensitivity
 of the derived parameters for M32 on the number of orbits in the
 library, etc.

In what follows, unless stated otherwise,
black hole masses are expressed in units of $10^6\msun$
and mass-to-light ratios in solar units in the $V$-band.

\section{Fits to Data Set A -- Constraining $\mh$ From Nuclear Data} 
\label{sec:pseudoA}

We first apply our modeling algorithm to various subsets of data set
A.  Data set A consists of kinematics within 1.5\arcsec, ``observed''
in such a way as to mimic observations of galactic nuclei with
HST/STIS.  In addition we include mass constraints out to 100\arcsec.
Figure~\ref{fig:pseudo-setA} shows the entire data set; the total
number of constraints is 571. No regularization (smoothing)
constraints were imposed in any of the models in this section.

In order to test the dependence of the modeling results on the number
and type of data points supplied to it, we defined restricted data
sets as follows:

a) A total of 98 constraints, consisting of the masses in 56 cells 
(every third radial cell and every third polar angle), 
and $v_l$ and $\sigma_l$ as measured in every third aperture.

b) A total of 164 constraints, consisting of the masses in 102 cells
(every other radial cell and every other polar angle),
and $v_l$ and $\sigma_l$ in 31 apertures.

c) A total of 226 constraints, consisting of the same mass constraints 
as in (b), as well as $v_l$, $\sigma_l$, and the GH moments $h_3-h_4$
measured at the same positions as in (b).

d) All 571 constraints, consisting of $19 \times 14$ cell masses,
and $v_l$, $\sigma_l$ and $h_3-h_4$ in all 61 apertures.

\noindent
We did not explicitly include the aperture masses shown  in
Figure~\ref{fig:pseudo-setA}b ($M^{\rm obs}_l$) in the fits (although
they are implicitly included as described in \S~\ref{sec:lpp}.)
However we verified that the aperture masses were always fitted to
better than 0.1\% for this data set.

\begin{figure}
\figurenum{2}
\label{fig:vary-const-fix-no}
\epsscale{1.}
\plotone{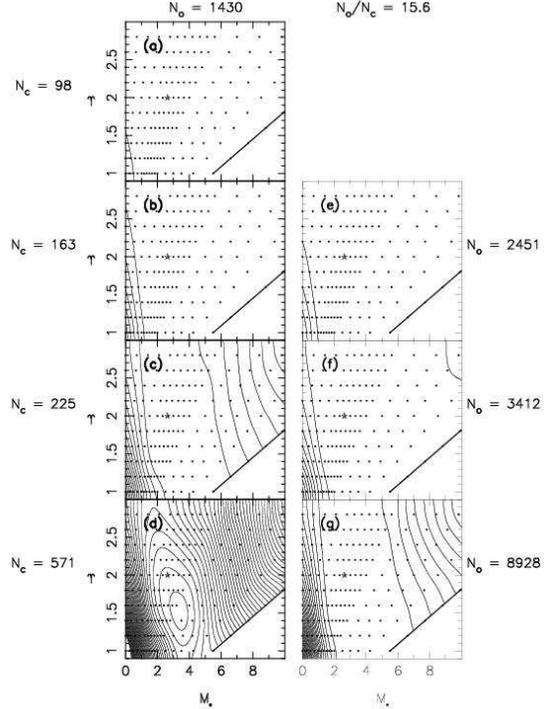}
\caption{Contour plots of $\chi^2(\mh,\Upsilon)$ for models
constructed to fit various subsets of data set A.
The star indicates the true model parameters.
Left column:
The number of orbits used in the solutions was fixed at $N_o=1430$.
(a) Fits to $v_l$ and $\sigma_l$ only, coarsely sampled; $N_c=98$.
(b) $v_l$ and $\sigma_l$ only, finely sampled; $N_c=163$.
(c) All four GH moments, finely sampled; $N_c=225$.
(d) All four GH moments, very finely sampled; $N_c=571$.
($N_c$ includes mass constraints.)
Right column:
Fits were carried out using the same data as in the left column, 
but now the number of orbits has been varied in order to
keep $N_o/N_c$ constant at 15.6.
(e) $N_o=2451$ (f) $N_o=3412$ (g) $N_o=8928$.
When the ratio of orbits to constraints is kept constant,
increasing the number of data points has little
effect on the tightness of the $\chi^2$ contours.
}
\end{figure}

The left column of Figure~\ref{fig:vary-const-fix-no} shows how the
$\chi^2$ contours change as the number of constraints is increased,
given a fixed number of orbits, $N_o=1430$.  It is clear that the
lowest velocity moments $v_l$ and $\sigma_l$ contain almost no
information about $\mh$ or $\Upsilon_V$: only when the higher GH
moments are added do the $\chi^2$ contours begin to exhibit a definite
minimum.  However the best-fit parameters in
Figure~\ref{fig:vary-const-fix-no}d are substantially displaced from
their true values and plots of the predicted kinematics confirm that
the fit to the data is poor.

\begin{figure}
\figurenum{3}
\label{fig:fix-Nc-vary-No}
\epsscale{0.75}
\plotone{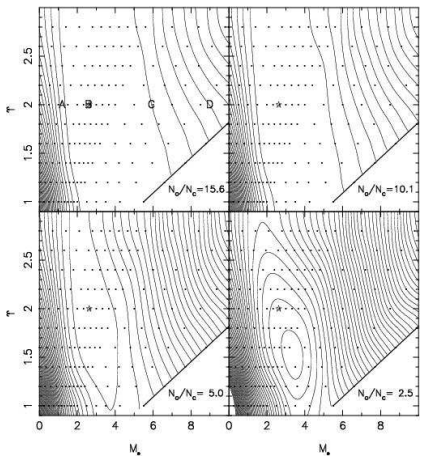}
\caption{Contour plots for a fixed set of observational constraints
(same as in Figure 2d, $N_c=571$) but different numbers of orbits,
as indicated.
The conclusions drawn from this data set about the
best-fit model parameters $\mh$ and $\Upsilon_V$ and their
uncertainties would depend very strongly on the number of orbits used
in the modeling.
The properties of the models labelled $A$- $D$ are illustrated
in Figure~\ref{fig:comp-mod-3dkin}.}
\end{figure}

A possible explanation for the offset and for the poor fit when the
number of data constraints is large, is the small ratio of orbits to
constraints in Figure~\ref{fig:vary-const-fix-no}d, $N_o/N_c = 2.5$.
This modest ratio -- while typical of the published modeling studies
(e.g. vdM98) -- suggests that our algorithm did not have much freedom
to explore different orbital solutions.  To test this idea, we
repeated the experiments but this time increased the number of orbits
in step with the number of constraints so as to keep the ratio
$N_o/N_c$ fixed.  The results are shown in the right panel of
Figure~\ref{fig:vary-const-fix-no}.  The differences are striking: we
now see that the topology of the first set of contours was an artifact
of the small number of orbits used.  When the number of orbits is
increased from 1430 to 8928 -- i.e.  when the ratio of orbits to
constraints is increased from 2.5 to ~15 -- the minimum in $\chi^2$
disappears, leaving only a broad $\chi^2$ plateau.  The true set of
model parameters lies within this plateau although there is no sense
in which this model can be said to be ``preferred.''  Evidently, even
the full set of GH moments can only weakly constrain the potential
when the modeling algorithm has the freedom to construct a wide
variety of orbital populations. It must be emphasized that in the
absence of smoothing constraints the actual number of orbits actually
used by the optimization routine is roughly equal to the total number
of constraints, irrespective of the size of the orbit
library. Increasing the size of the orbit library basically increases
the availability of orbits with the right kind of properties in
the right part of phase space.

\begin{figure}
\figurenum{4}
\label{fig:kin-orbratio}
\epsscale{1.}
\plotone{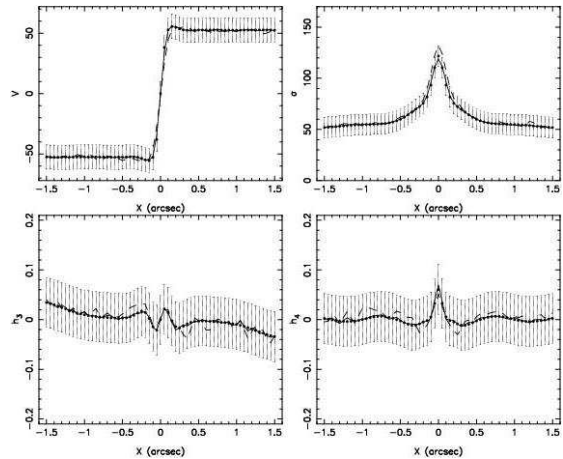}
\caption{Fits to the kinematical data (Fig. 1) for two 
orbital solutions that lie
within the $\chi^2$ valleys of Fig. 3, 
close to the true model ($\star$). 
Solid line: $N_o/N_c= 5$; dashed line: $N_o/N_c=2.5$. 
Models constructed using the two larger values of $N_o/N_c$ 
shown in Fig. 3 provide almost perfect fits to these data;
those fits are not shown here.}
\end{figure}

In these experiments, the number of observational constraints was
varied. More typically one is faced with a fixed number of
measurements. Figure~\ref{fig:fix-Nc-vary-No} shows what happens when
$N_c$ is fixed -- we used the full data set A, with $N_c=571$ -- but
the number of orbits is varied.  Again we see that the topology of the
$\chi^2$ plot depends strongly on the ratio of orbits to constraints.
As $N_o/N_c$ increases from 2.5 to 5.0, the $\chi^2$ contours shift so
that their apparent center is close to the true model parameters, but
as $N_o/N_c$ is increased still more, all semblance of a unique
$\chi^2$ minimum vanishes and the potential parameters become
essentially unconstrained.  Indeed it is not clear from these plots
whether we have reached a limit; the $\chi^2$ valley may become even
broader as $N_o/N_c$ is increased above 15.6.  In the plots with the
two largest values of $N_o/N_c$, models lying within the $\chi^2$
plateau provide essentially perfect fits to the kinematical data and
each of the mass constraints is fit to better than one part in $10^6$.
Figure~\ref{fig:kin-orbratio} shows the quality of the fit to the data
in the cases $N_o/N_c=5.0$ and $2.5$; the most significant deviations
are in $h_4$.

\begin{figure}
\figurenum{5}
\label{fig:chi1D-GHA}
\epsscale{0.8}
\plotone{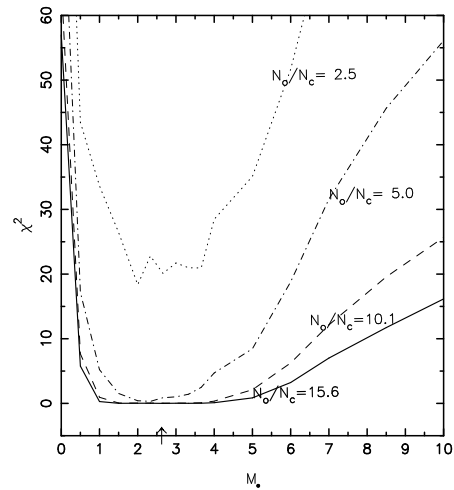}
\caption{1-D cuts through the
$\chi^2$ plots of Fig.~\ref{fig:fix-Nc-vary-No}, all taken at
$\Upsilon_V = 2.$ The vertical arrow indicates the location of the
true model parameter, $\mh = 2.625\times10^6\msun$.
When the number of orbits used is small, there is a definite,
but spurious, $\chi^2$ minimum.
As $N_o$ is increased, this minimum broadens into the perfectly flat
plateau characteristic of under-determined problems.
The true model parameters lie on that plateau but can not be
unambiguously recovered.}
\end{figure}

Figure~\ref{fig:chi1D-GHA} shows 1D cuts through the $\chi^2$ plots of
Figure~\ref{fig:fix-Nc-vary-No}, all taken at $\Upsilon_V = 2.$ As the
ratio $N_o/N_c$ increases, two things happen: the absolute value of
$\chi^2$ drops, reflecting the better quality of the fit as the number
of orbits is increased; and the $\chi^2$-valley becomes broader.  The
plateau of precisely-constant $\chi^2$ predicted in \S 3 is very clear
for $N_o/N_c\gtrsim 5$.  The true value of $\mh$ lies within this
plateau but there is no sense in which it is preferred.  This behavior
of the $\chi^2$ plots as $N_o$ is varied was first predicted by
Merritt \& Ferrarese (2001) (their Fig. 7).

\begin{figure}
\figurenum{6}
\epsscale{0.60}
\plotone{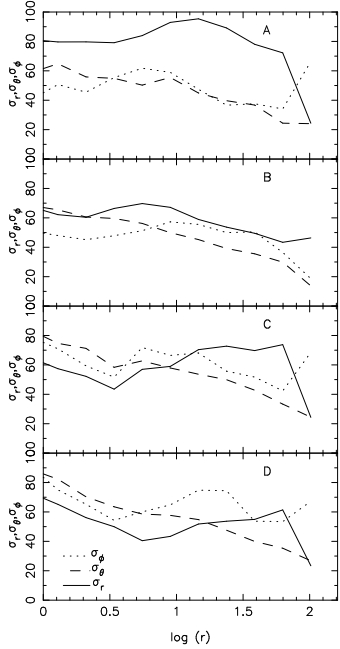}
\label{fig:comp-mod-3dkin}
\caption{The intrinsic velocity dispersions $\sigma_r, \sigma_{\phi},
\sigma_{\theta}$ as functions of radius for models A-D in
Fig.~\ref{fig:fix-Nc-vary-No}. All models have comparable $\chi^2$
values and $\Upsilon_V = 2$. 
Black hole masses are: A, $1\times
10^6\msun$; B, $2.66\times 10^6\msun$; C, $6.\times
10^6\msun$; D, $8.5\times 10^6\msun$. 
The values of $\Upsilon$ and $\mh$ used in constructing Model B are 
closest to the true values.
This model is approximately isotropic ($\sigma_r\approx\sigma_{\theta}$),
as was the 2I model from which the data were generated.}
\end{figure}

The internal velocity dispersions in four models (labeled A-D in
Figure~\ref{fig:fix-Nc-vary-No}a) are shown in
Figure~\ref{fig:comp-mod-3dkin}.  The models all have $\chi^2$ values
comparable to the model with the true potential parameters.  Close to
the center, the model with lowest $\mh$ (A) has a significantly larger
number of stars on radial orbits than the models with large $\mh$
(C-D); the increase in $\sigma_r$ is needed to keep the central
velocities high in spite of a too-small black hole.  Nevertheless, so
great is the freedom to choose different orbital populations that even
knowledge of the projected GH moments can not rule out these extreme
models.

\begin{figure}
\figurenum{7}
\epsscale{0.9}
\plotone{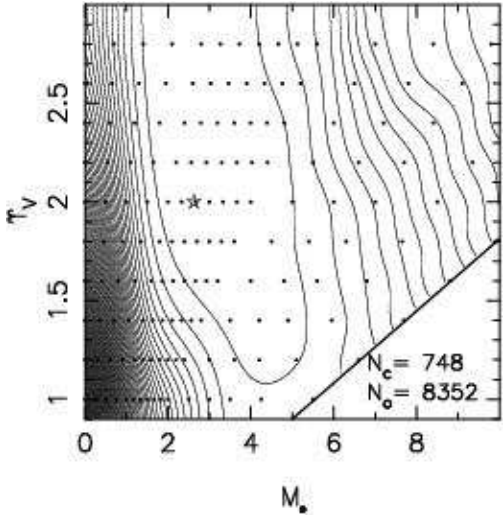}
\label{fig:stswht} 
\caption{$\chi^2$ contours for fits to the data from data set A, to which
has been added the simulated data from the ``WHT'' ground-based 
apertures.}
\end{figure}

It is essential to point out that part of the indeterminacy
illustrated in Figure~\ref{fig:fix-Nc-vary-No} might be due to the
fact that the data of data set A are restricted to the region near the
black hole; hence the model kinematics are not at all constrained at
large radii. This means that the modeling algorithm has unlimited
freedom to vary the properties of the model at large radii while
fitting the small radius data. In order to test if this is the sole
cause of the indeterminacy we show in Figure~\ref{fig:stswht} how the
$\chi^2$ contours are modified if, in addition to data set A
(kinematical data extending to 1.5\arcsec), the modeling algorithm is
given the additional $44$ data points (including the first 4 moments
of the LOSVD at each point) from data set B that correspond to the
ground-based WHT observations along all position angles (kinematical
data extending out to 8\arcsec). We see once again that when the full
orbit library of $\sim 9000$ orbits is used a long flat $\chi^2$
valley which is somewhat more restricted in $\mh$ results.

As an alternative to fitting GH moments, one can fit directly to the
LOSVDs from which the GH moments were derived (e.g. \cite{Merritt97}).
This procedure is expected to be inefficient if the LOSVDs are nearly
Gaussian since measurements at many distinct velocities are required
to reproduce accurate estimates of just the lowest-order GH moments.
But direct use of the LOSVDs may be advisable near the centers of
galaxies where observations can reveal extended wings due to
high-velocity motion around the black hole (e.g. \cite{Joseph01}),
wings that are poorly represented by the lowest terms in a GH
expansion.

\begin{figure}
\figurenum{8}
\epsscale{0.9}
\plotone{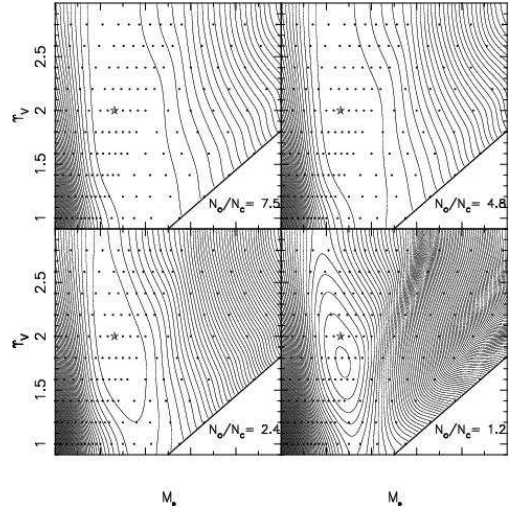}
\label{fig:compOR-los}
\caption{$\chi^2$ contours for fits to the full LOSVDs in all 61
apertures of data set A; $N_c = 1198$.  $N_o = 8928, 5775, 2863,1430$ 
in a-d respectively.}
\end{figure}

Figure~\ref{fig:compOR-los} shows $\chi^2$ contours for fits to the
full LOSVDs, sampled at $\Delta v \sim 40$\kms.  This velocity
sampling is approximately equal to the velocity resolution of the STIS
spectrograph at 8500\AA. (The velocity scale of the the STIS
spectrograph at $\sim 8500$\AA\, is $\sim 19$\kms per pixel. Thus two
pixels in the spectral direction (Nyquist sampling) imply a velocity
resolution of $\sim 38$\kms~). A more pragmatic justification is that
sampling at $\Delta v \sim 40$\kms\ already implies 1198 constraints
and halving the velocity spacing would increase the number of
constraints to over 1800, requiring a prohibitively large number of
orbits for the modeling.  We carried out optimizations for the same
four sets of orbits ($N_o = 8928, 5775, 2863, 1430$) used to fit the
GH moments in Figure~\ref{fig:fix-Nc-vary-No}.  The total number of
data constraints was 1198: the same set of 266 mass constraints as in
Figure~\ref{fig:fix-Nc-vary-No}, and the LOSVDs measured at all 61
apertures along the major axis.  The ratio $N_o/N_c$ is smaller than
in the plots of Figure~\ref{fig:fix-Nc-vary-No} because of the roughly
three times larger number of constraints required to represent the
LOSVDs.

\begin{figure}
\figurenum{9}
\epsscale{1.}
\plotone{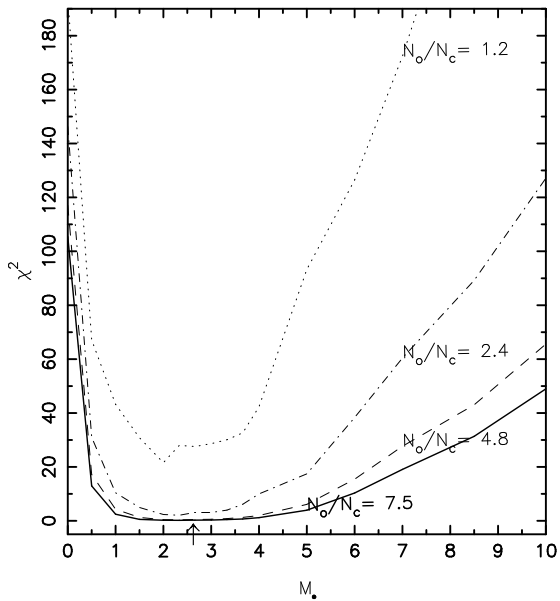}
\label{fig:chi1d-LOSA}
\caption{1-D cuts through
Figure~\ref{fig:compOR-los} for $\Upsilon_V=2$.
The unique minimum in $\chi^2$ that appears when the number
of orbits is small, becomes a perfectly flat plateau when 
$N_o$ is large, indicating that the estimation of $\mh$ from these
data is under-determined.} 
\end{figure}

In all four panels of Figure~\ref{fig:compOR-los}, the decrease in
$N_o/N_c$ relative to Figure~\ref{fig:fix-Nc-vary-No} results in a
slightly smaller allowed range of models.  But once again, for a large
enough orbit library, there is an extended region within which
$\chi^2$ is precisely constant.  For the smallest orbit library ($N_c
= 1430$) the true solution lies outside
the minimum contour and the ``best fit'' solution is obtained for a
larger $\mh$ and smaller $\Upsilon$ than those corresponding to the
true solution.  Figure~\ref{fig:chi1d-LOSA} shows 1D cuts through
Figure~\ref{fig:compOR-los} for $\Upsilon_V=2$.  The constant-$\chi^2$
plateau appears for $N_o/N_c\gtrsim 5$.

In order to make a more reasonable comparison between the quality of
the fits to the LOSVDs and to the GH moments, we defined a new
statistic $\chi^2_{\rm kin}$, which measures only the goodness of fit
to the kinematical data in each aperture i.e. $V$, $\sigma$, $h_3$ and
$h_4$, rather than the value of the objective function (which in this
case includes the LOSVDs). (The $\chi^2$ of the fit to the mass
constraints is also excluded from $\chi^2_{\rm kin}$ but is $<
10^{-3}$ everywhere).  When 8928 orbits were used, fitting to the
LOSVDs gave a minimum $\chi^2_{\rm kin} = 0.416$, while fitting to the
GH moments gave $\chi^2_{\rm kin} = 0.0442$.  (Although there is
nearly an order of magnitude difference in the two numbers, the two
fits are indistinguishable to the eye and both are virtually perfect.)
Thus we conclude that fitting to the GH moments may be adequate even
when the LOSVDs have large wings, as in the case of our central
aperture.

\begin{figure}
\figurenum{10}
\epsscale{1.}
\plotone{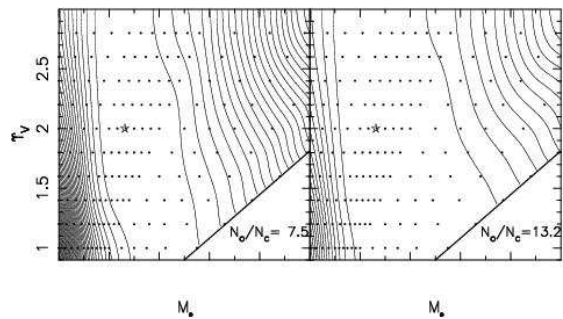}
\label{fig:complos-delv}
\caption{Fits to the LOSVDs sampled with $\Delta v= 40$ \kms\ 
(left panel) and $\Delta v= 100$ \kms (right panel)
at all 61 apertures using the full library of 8928 orbits.}
\end{figure}

\begin{figure}
\figurenum{11}
\epsscale{1.}
\plotone{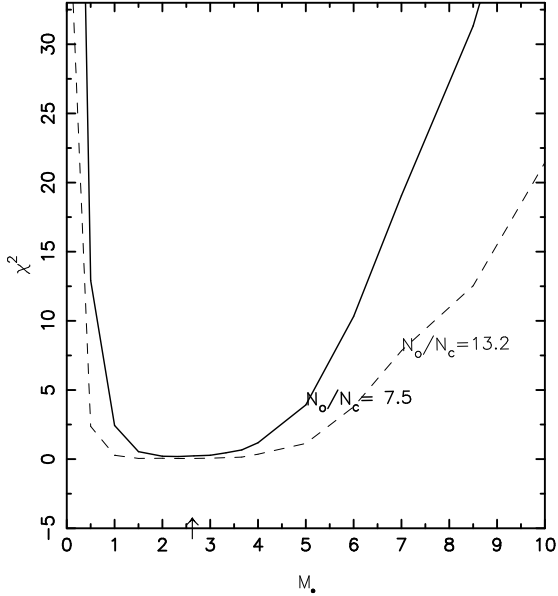}
\label{fig:chi1d-delv}
\caption{1-D cuts through Figure~\ref{fig:complos-delv} at
$\Upsilon_V = 2$. 
The solid line is for $\Delta v = 40$ \kms\ and the
dashed line is for $\Delta v = 100$ \kms.
Arrow marks true value of $\mh$.}
\end{figure}

Prior to the installation of STIS aboard HST, the faint object
spectrograph (FOS) was used to observe the nuclei of galaxies with
high spatial resolution, although its velocity resolution was only
$\sim 100$\kms. Due to the difficulties associated with reducing the
FOS data, only a few of the galaxies observed with the FOS have been
modeled.  These include M32 (vdM98) and NGC 3379 \cite{Gebhardt00a}.
vdM98 used $V_l$ and $\sigma_l$ as derived from
the FOS observations in their modeling of M32, while Gebhardt et
al. (2000a) attempted to extract the central few LOSVDs in NGC~3379,
sampled at 100 \kms\ spacing.  In their most recent paper Gebhardt et
al. (2003) modeled the kinematics of 12 galaxies with nuclear data
from STIS.  In all cases they sample the LOSVDS with only 13 points
with typical velocity spacing of $\sim 100$\kms.  In
Figure~\ref{fig:complos-delv} we compare the fits to LOSVDs sampled at
40~\kms\ and 100~\kms\ at all 61 apertures using the full orbit
library of 8928 orbits. This plot shows that when LOSVDs are coarsely
sampled with $\Delta v = 100$\kms, a much larger region of parameter
space can fit the data equally well and the model parameters are not
well constrained.  Figure~\ref{fig:chi1d-delv} shows 1D cuts through
Figure~\ref{fig:complos-delv} at $\Upsilon_V = 2$.  For the model
closest to the ``true'' model ($\mh = 2.66$, $\Upsilon_V = 2$),
$\chi^2 = 0.416$ and $\chi^2 =0.084$ for $\Delta v = 40$\kms\ and
$\Delta v = 100$\kms\ respectively.

\begin{figure}
\figurenum{12}
\epsscale{1.}
\plotone{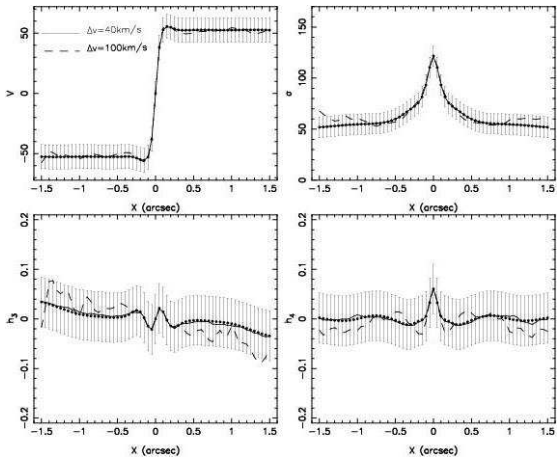}
\label{fig:kin-delv}
\caption{Fits to the kinematical data ($V_l, \sigma, h_3, h_4$)) for
models from Fig. 10 with $\mh = 2.66$, $\Upsilon_V = 2$. 
The solid line is the fit
obtained with $\Delta v = 40$\kms ($\chi^2_{\rm kin} = 4.16\times
10^{-1}$) and the dashed line is the fit obtained with $\Delta v =
100$\kms ($\chi^2_{\rm kin} = 32.5 $).}
\end{figure}

\begin{figure}
\figurenum{13}
\epsscale{1.}
\plotone{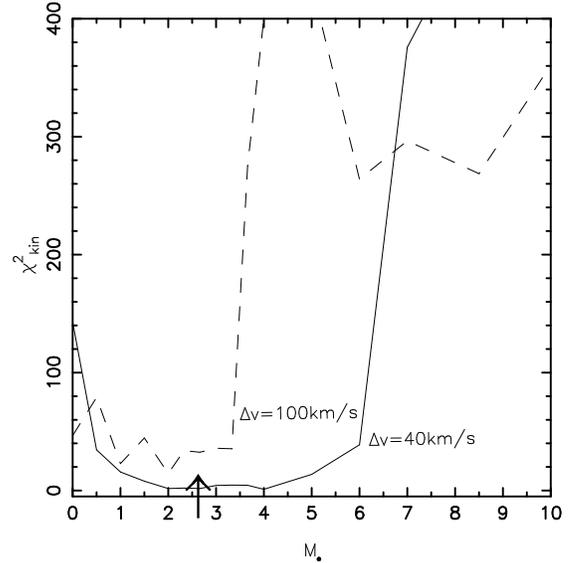}
\label{fig:chi1d-kin-delv}
\caption{1-D plot of $\chi^2_{\rm kin}$ for $\Upsilon_V = 2$ for fits
with the two different velocity spacings ($\Delta v = 40$\kms and
$\Delta v = 100$\kms).}
\end{figure}

From these $\chi^2$ values, one might conclude that all models close
to the bottom of the $\chi^2$ valley would give equally good fits.
However, it is once again essential to compare how the kinematics
would be fitted if all the information in the best sampled LOSVDs were
used.  To do this we use the orbital weights provided by the fits to
the LOSVDs sampled at 40~\kms\ and 100~\kms\ but co-add the
appropriately weighted GH moments computed from the orbital LOSVDs
sampled at 5~\kms. Figure~\ref{fig:kin-delv} shows the fits the GH
moments for models lying on the plateau of the $\chi^2$ valley with
each of the two velocity spacings.  It is clear that fitting
coarsely-sampled LOSVDs gives a much poorer fit to the kinematical
data, especially for the higher-order GH moments, e.g. $h_4$. This is
despite the fact that they are an almost perfect fit to the coarsely
sampled LOSVDs! This quality of the fit worsens even more at points
further away from the true model as shown by the steeply rising and
highly variable $\chi^2_{\rm kin}$ values plotted in
Figure~\ref{fig:chi1d-kin-delv}. This is understandable since both $V$
and $\sigma$ at large radii are $\sim 50$\kms roughly half the spacing
between points in the LOSVD!

The suitability of LOSVDs sampled at $\sim 100$\kms is likely to be
particularly bad in compact low luminosity ellipticals like M32 where
the central velocity dispersion is $\leq 150$~\kms but may be less
problematic in large giant ellipticals where the central velocity
dispersion is $\sim 250-300$~\kms. It is clear however that using a
fixed number of grid points per LOSVD for all galaxies could produce
non-uniform results. This implies that it is essential to tailor the
modeling parameters to each galaxy.

\section{Fits to Data Set B -- A 2I Model of M32} 

\begin{figure*}
\figurenum{14}
\epsscale{1.}
\plotone{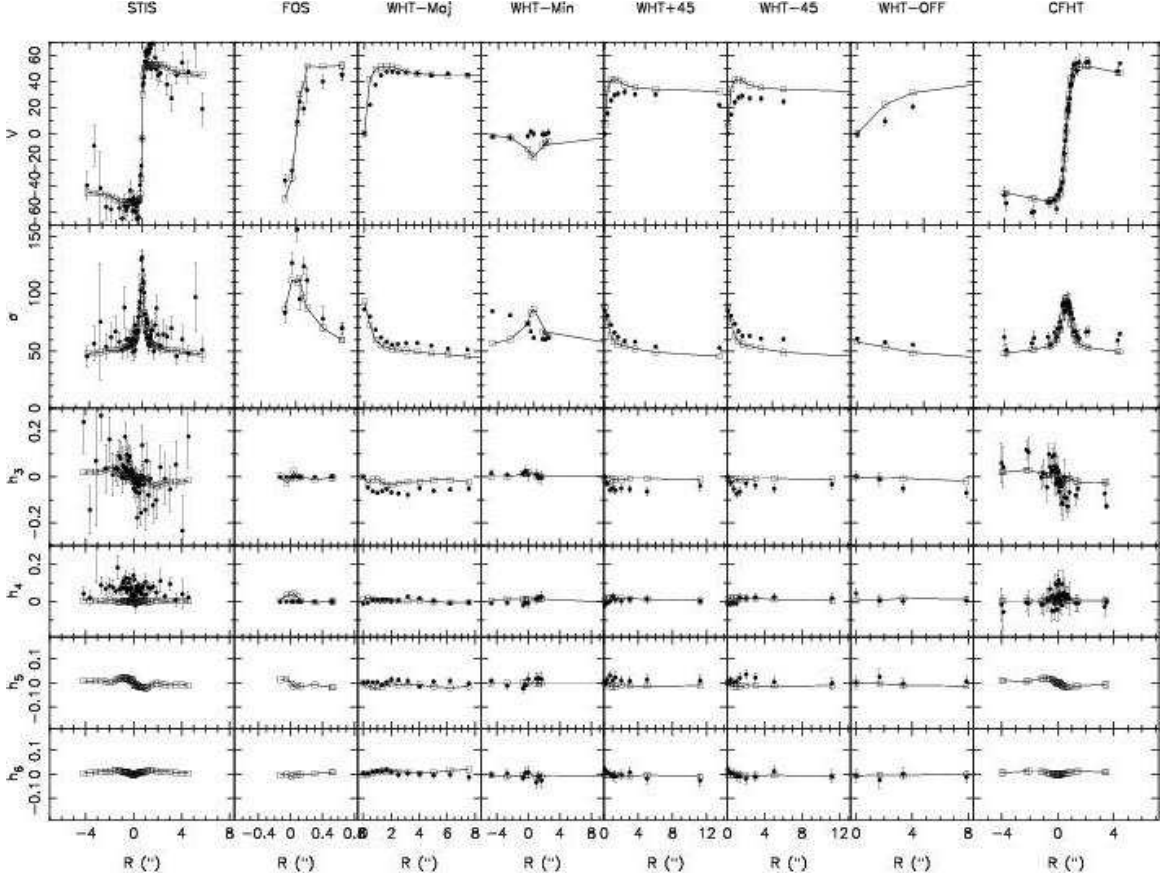}
\label{fig:PseudodatB}
\caption{Data set B.  Solid dots represent the real data to
which the 2I model was fitted; solid lines represent the
2I fit; open squares represent the points on the fit which
were selected as data set B.}
\end{figure*}

Data set B was obtained by ``observing'' the 2I model through exactly
the same set of apertures, and with the same PSFs, as in the
observations of M32 (van der Marel et al.~1994a; Bender et al.~1996; vdM98;
Joseph et al.~2001) that were used to construct the 2I model described
in section \S~3. vdM98 used this same set of
observations in building their 3I models of M32 and estimating the
black hole mass.  Figure~\ref{fig:PseudodatB} shows that data set B is
not a perfect match to the actual M32 data although it reproduces the
kinematics near the central black hole very well. Error bars on the
pseudo dataset were defined as described in \S~\ref{sec:2I_model}.

\begin{figure}
\figurenum{15}
\epsscale{0.75}
\plotone{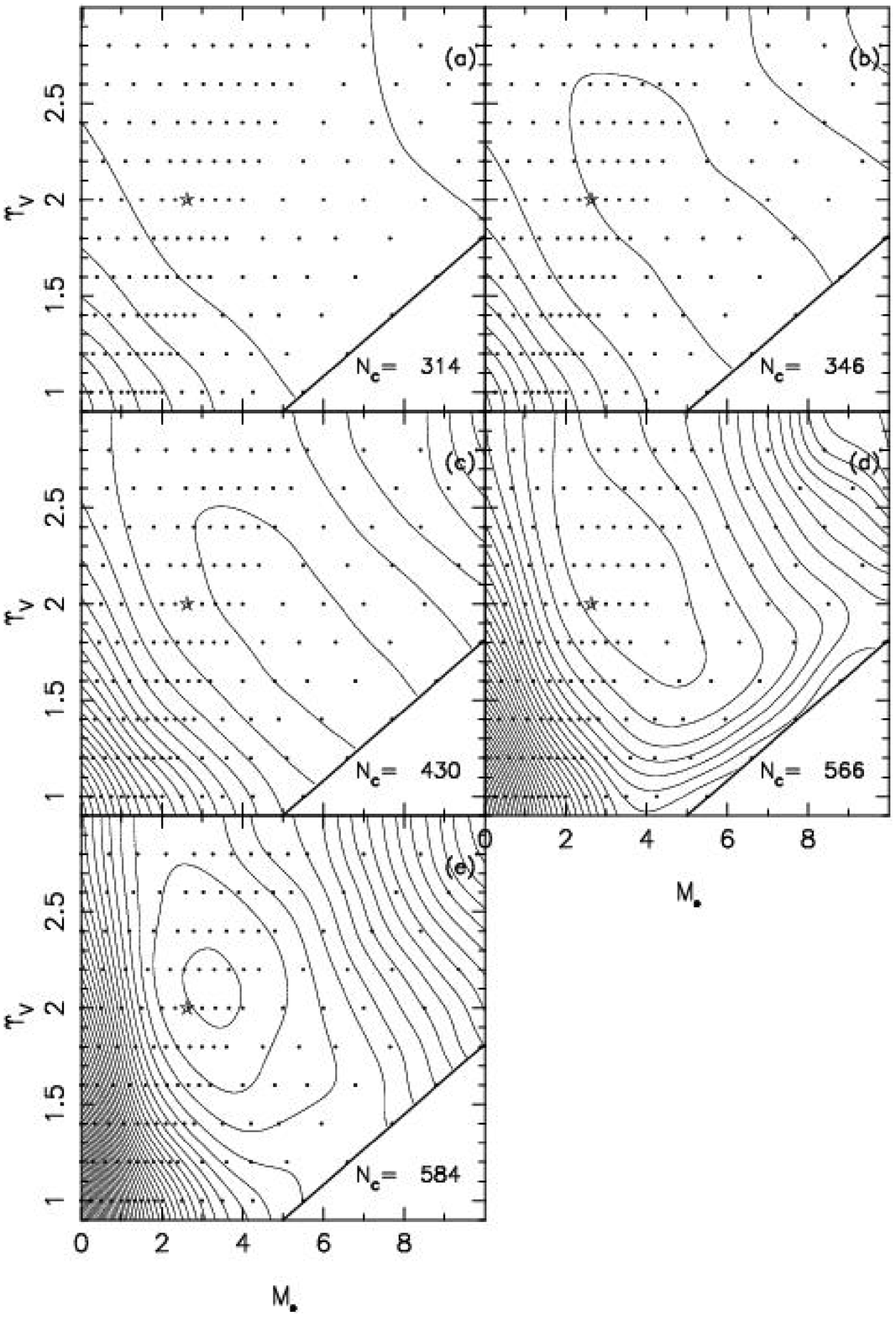}
\label{fig:cont-vdm1ax}
\caption{ Contour plots of the $\chi^2$ that measures
the quality of the fit to various subsets of data set B
(the simulated M32 data) using 1982 orbits. 
(a) Major axis $V$ and $\sigma$ (WHT,
CFHT apertures); (b) major axis $V$, $\sigma$, $h_3$, $h_4$ 
(WHT, CFHT apertures); (c) major and minor axis $V$, $\sigma$, 
$h_3$, $h_4$ (WHT, CFHT apertures); 
(d) $V$, $\sigma$, $h_3$, $h_4$ along all PAs (WHT, CFHT apertures); 
(e) all of the constraints in (d) plus $V$ and $\sigma$ from the
FOS apertures. 
The $\star$ labels the true model parameters.
$N_c$ is the total number of constraints including mass constraints.
This plot seems to suggest that the constraints on $\mh$ and
$\Upsilon$ become rapidly tighter as the number of data points 
increases.}
\end{figure}

In Figure~\ref{fig:cont-vdm1ax}, we repeat an experiment first carried
out by vdM98 in their analysis of the actual M32 data (see their
Appendix A).  We fixed the number of orbits in our 3I modeling
algorithm at $N_o=1982$ -- similar to the number (1960) used by those
authors -- and explored how the $\chi^2$ contours change as we apply
progressively larger numbers of observational constraints, as follows
(all from data set B): (a) major axis $V$ and $\sigma$ in the WHT and
CFHT apertures; (b) major axis $V$, $\sigma$, $h_3$, $h_4$ (WHT,
CFHT); (c) major and minor axis $V$, $\sigma$, $h_3$, $h_4$ (WHT,
CFHT); (d) $V$, $\sigma$, $h_3$ and $h_4$ along all position angles
(WHT, CFHT); (e) all constraints in (d) plus $V$ and $\sigma$ from the
HST/FOS apertures.  Each of these fits included 266 meridional-plane
mass constraints within 100\arcsec.

As in the previous section none of the models discussed in this
section were constructed with regularization constraints imposed.

\begin{figure}
\figurenum{16}
\epsscale{1.}
\plotone{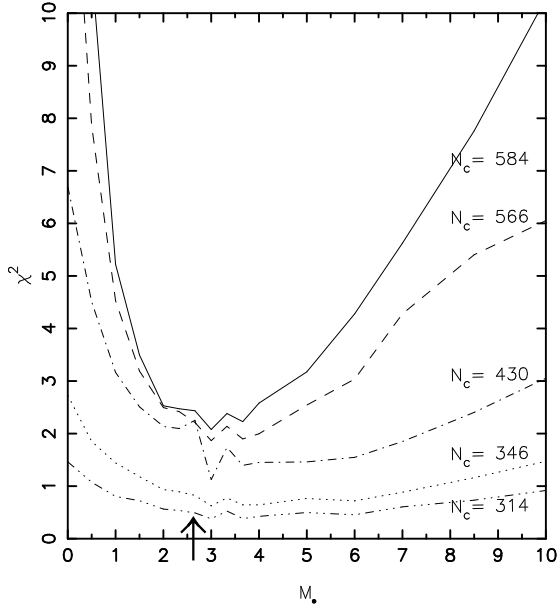}
\label{fig:vdm1ax-1D}
\caption{1-D cuts through Figure~\ref{fig:cont-vdm1ax} at
$\Upsilon_V=2$. Arrow indicates true value of $\mh$.}
\end{figure}

Figures~\ref{fig:cont-vdm1ax} and \ref{fig:vdm1ax-1D} show how the
constraints on $\mh$ and $\Upsilon$ appear to tighten as the number of
data points provided to the modeling algorithm are increased.  When
only the major-axis ``WHT'' measurements of $V$ and $\sigma$ are used,
the potential parameters are almost unconstrained, but when the entire
data set is given to the modeling algorithm, a well-defined minimum
appears in $\chi^2(\mh,\Upsilon)$ that is reasonably close to the true
model parameters.  vdM98 found a similar dependence of the $\chi^2$
contours on number of data points when modeling the true M32 data.

\begin{figure}
\figurenum{17}
\epsscale{0.85}
\plotone{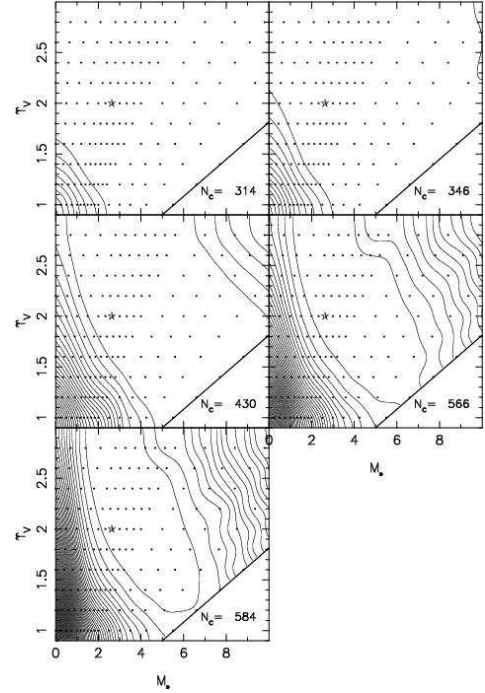}
\label{fig:cont-vdm2ax}
\caption{Same as Figure~\ref{fig:cont-vdm1ax}, except that the
size of the orbit library in each panel has been adjusted such that 
$N_{o}/N_c$ is constant at $\sim 10$.
The $\chi^2$ contours now change much more slowly as the number of
data points is increased, and even for the full data set,
the constraints on $\mh$ and $\Upsilon$ are weak.}
\end{figure}

\begin{figure}
\figurenum{18}
\epsscale{1.}
\plotone{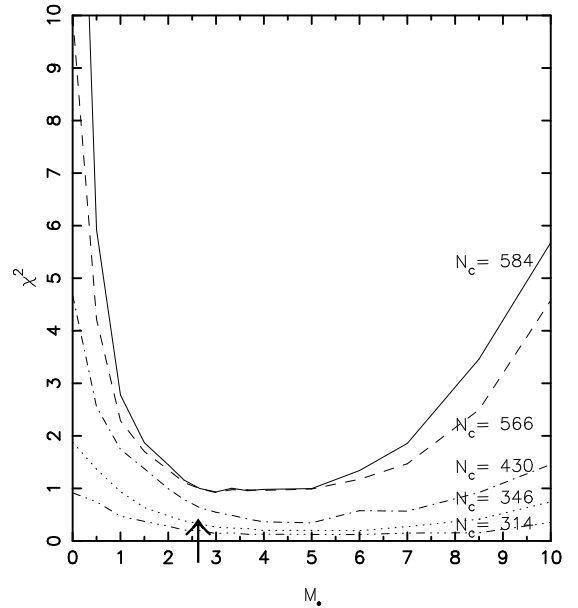}
\label{fig:vdm2ax-1D}
\caption{1-D cuts through Figure~\ref{fig:cont-vdm2ax} at
$\Upsilon_V=2$.
Even using the full data set ($N_c=584$),
there is a plateau of constant $\chi^2$ indicating that
these simulated data do not uniquely constrain $\mh$.}
\end{figure}

But Figures~\ref{fig:cont-vdm2ax} and \ref{fig:vdm2ax-1D} tell a very
different story.  Now the fits have been carried out using a {\it
fixed} ratio of orbits to data constraints, $N_o/N_c \approx 10$.  The
rapid shrinking of the $\chi^2$ contours with increasing $N_c$ in
Figure~\ref{fig:cont-vdm1ax} and \ref{fig:vdm1ax-1D} is now gone: even
using the full set of data gives a $\chi^2(\mh)$ plot with an extended
flat plateau, stretching from $\mh\approx 2\times 10^6\msun$ to
$\mh\approx 6\times 10^6\msun$.  The true value, $\mh=2.625\times
10^6\msun$, lies on the edge of this plateau suggesting that even the
large number of orbits we used (5856) is barely sufficient to
reproduce the true $\chi^2$ contours.

The most important conclusion we draw from a comparison of
Figures~\ref{fig:cont-vdm1ax} and \ref{fig:cont-vdm2ax} is that {\it
the appearance of the $\chi^2$ contours depends strongly on the
flexibility of the modeling algorithm}.  The quality of the fit to the
data depends at least as strongly on the size of the orbit library as
on the size of the data set.  Comparisons between fits made with
different sets of data are problematic unless care is taken to
demonstrate that the ratio $N_o/N_c$ is sufficiently large for each
fit.  And for a given data set, statements about the best-fit model
parameters and their confidence intervals can be very strongly
influenced by the number of orbits used.

We note that including the ``HST/FOS'' measurements from data set B
has almost no influence on the range of indeterminacy in $\mh$; the
width of the constant-$\chi^2$ plateau is virtually unchanged
(Figure~\ref{fig:vdm2ax-1D}).  This suggests that the FOS data for M32
did not significantly tighten the constraints on the mass of the black
hole in this galaxy compared with the constraints set by the
ground-based data.  vdM98 reached a different
conclusion; comparison of Figures~\ref{fig:cont-vdm1ax} and
\ref{fig:cont-vdm2ax} suggests that they may have been misled by the
relatively small number of orbits in their modeling algorithm.

\begin{figure}
\figurenum{19}
\epsscale{1.}
\plotone{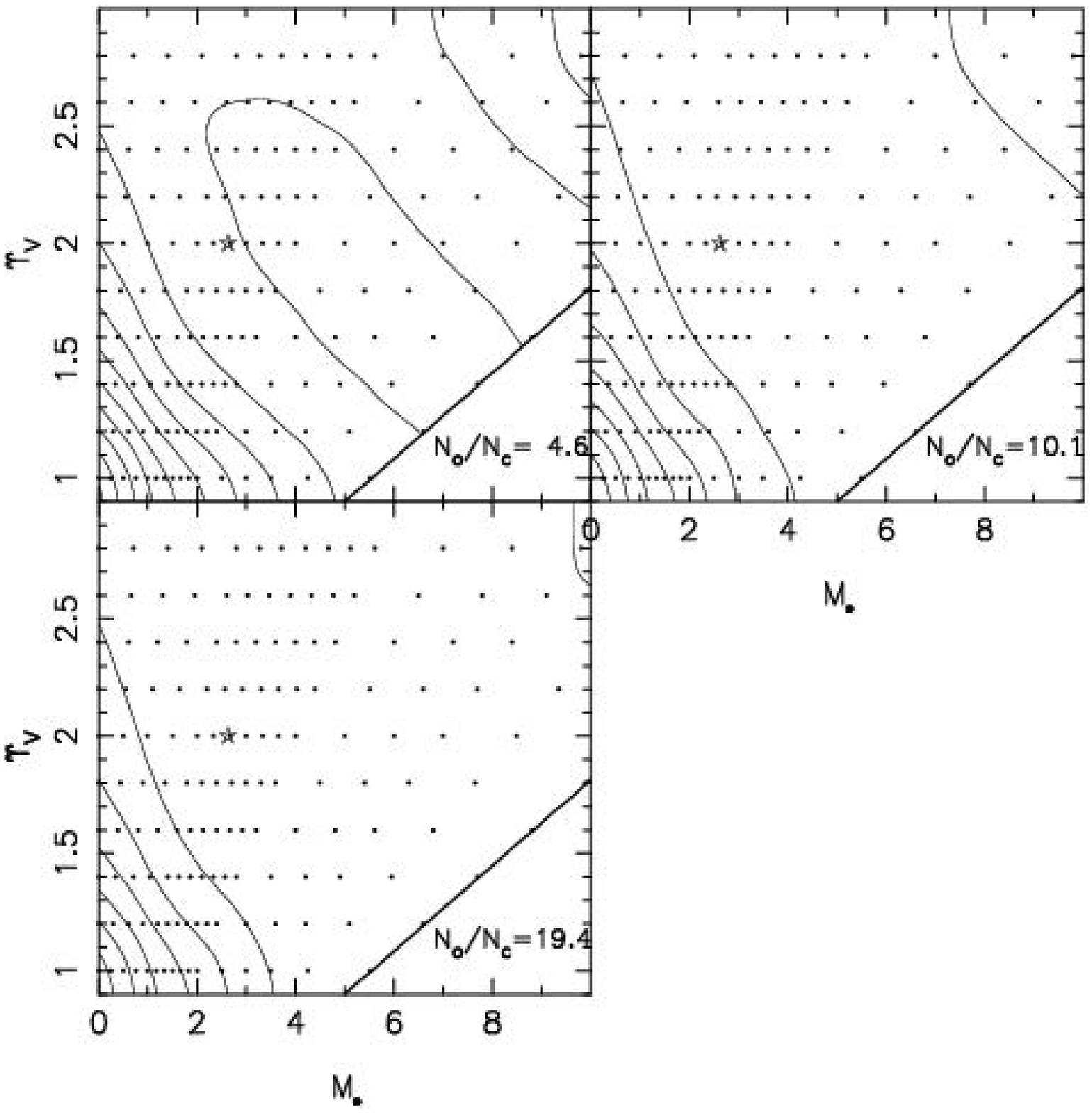}
\label{fig:vdm3_NobyNc}
\caption{Fits to the subset of data set B corresponding to the
ground-based, WHT apertures only, for various numbers of orbits;
$N_c=430$.}
\end{figure}

\begin{figure}
\figurenum{20}
\epsscale{1.}
\plotone{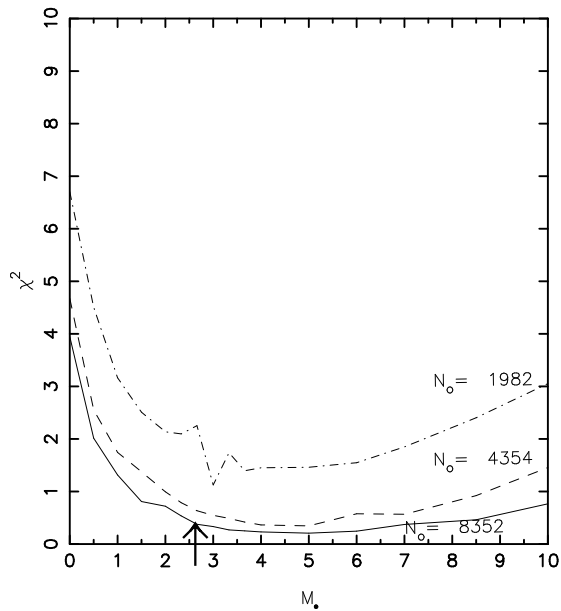}
\label{fig:vdm3_NobyNc1d}
\caption{1-D cuts through Figure~\ref{fig:vdm3_NobyNc} at
$\Upsilon_V=2$.
These data, which are superior in quality to most
STIS/HST nuclear data, place only very weak constraints
on $\mh$.}
\end{figure}

It is interesting to compare these results with those obtained using
only the subset of data set B corresponding to the ground-based, WHT
data; these (simulated) data have an effective resolution ${\rm
FWHM}/2r_h\approx 0.5$, better than that of most galaxies observed
with HST/STIS (\cite{MerrittF01b}; \cite{Gebhardt02}) and their
spatial coverage and S/N are much greater than those of most STIS
nucleus data.  Thus we extracted from data set B measurements at all
the WHT apertures of $V$, $\sigma$, $h_3$ and $h_4$, including all
position angles (430 constraints).  Figures~\ref{fig:vdm3_NobyNc} and
\ref{fig:vdm3_NobyNc1d} show the results, for three different numbers
of orbits, $N_o = (1982, 5674, 8352)$.  When the ratio of orbits to
constraints is largest ($N_o/N_c = 19.4$ for $N_o=8352$), excellent
fits are obtained for {\it any} black hole mass in the range $1\times
10^6\msun\lesssim\mh\lesssim 10\times 10^6\msun$!  While there is a
hint of a minimum at $\mh\approx 4.5\times 10^6\msun$, it is well
removed from the true value of $\mh$ and furthermore its location is
very sensitive to $N_o$.  We conclude that these data are almost
useless for constraining the black hole mass.  We would expect a
similar or greater degree of indeterminacy in values of $\mh$ derived
from many of the galactic nuclei observed with HST/STIS.  We return to
this point in \S~8.

\begin{figure}
\figurenum{21}
\epsscale{1.}
\plotone{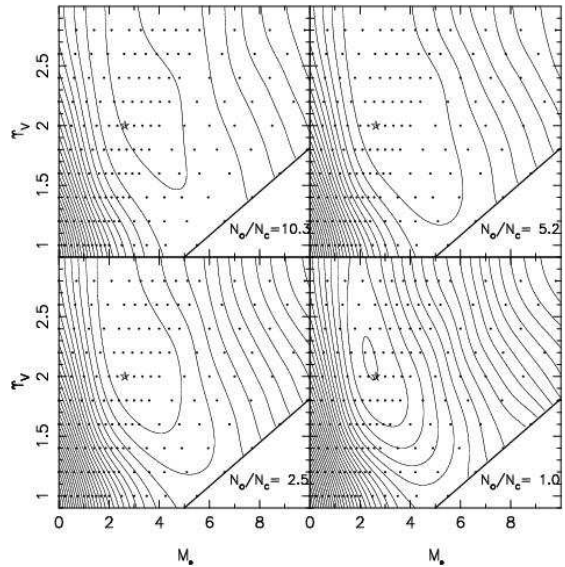}
\label{fig:orbratio-B140}
\caption{Contour plots of $\chi^2$ for models constructed to fit data
set B including HST/STIS apertures as well.  The same number of
constraints (810) are fitted in each panel but the number of orbits in
the library is varied as indicated on the plots.}
\end{figure}

\begin{figure}
\figurenum{22}
\epsscale{1.} 
\plotone{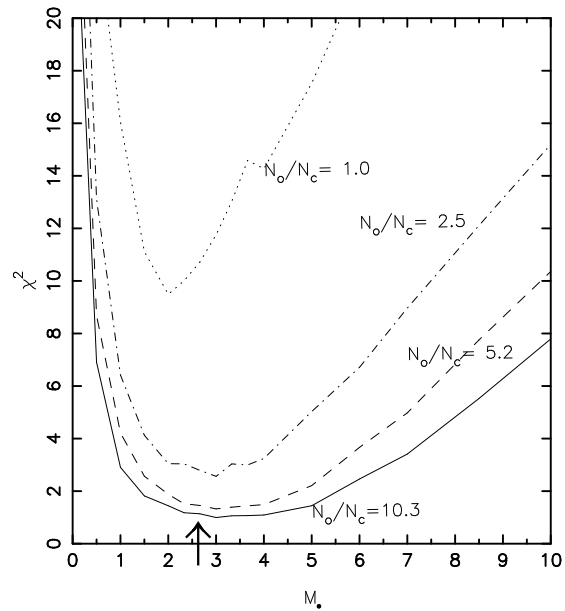}
\label{fig:chi1D-B140}
\caption{1-D cuts through Fig.~\ref{fig:orbratio-B140}. This figure
shows that for the largest orbit library, the minimum in the $\chi^2$
valley is still reasonably narrow, suggesting that the HST/STIS data
for M32 may yield tight constraints on $\mh$ in this galaxy.}
\end{figure}

Finally we ask if the constraints on $\mh$ using this data set can be
narrowed by adding the simulated HST/STIS data.
Figure~\ref{fig:orbratio-B140} shows $\chi^2$ contours for 3I model
fits to data set B including all the STIS apertures as well as the
ground based and FOS apertures.  The data were fitted at 140 apertures
in total (the four outermost STIS data points on one side of the
galaxy were the only apertures in the Joseph et al. (2001) data set
that were not fitted).  Figure~\ref{fig:chi1D-B140} shows 1-D cuts
through the $\chi^2$-contour plot at $\Upsilon = 2$.  This figure
suggests that the addition of the STIS data to the existing data for
M32 may yield a tight constraint on $\mh$: even for the largest orbit
library, the allowed range of solutions is quite small.  We note also
that the true solution lies close to the center of the minimum in the
$\chi^2$ valley.

Figure~\ref{fig:orbratio-B140}(a) should be considered provisional
since the ratio of orbits to constraints is $\sim 10$ and likely to be
marginally adequate.  We will return to this point in a later paper
when we analyze the observed STIS data for M32.

\section{Data Set C: M32 Re-Examined}

In \S~5 and 6 we presented $\chi^2$-plots of fits to two simulated
data sets derived from a model that was based on data from M32.  Here
we show the results of fits to the actual data used in the
construction of that model, our data set C.  These are the same data
used by vdM98 in their 3I study of M32.  The
constraints in our data set C include meridional plane masses in 266
cells. In the modeling results presented below, mass constraints were
fit to an accuracy of better than 3\% everywhere and to better than
0.1\% at all points within the minimum $\chi^2$-valley.

The models discussed in this section were constructed without
regularization constraints.  The same is true of the vdM98 modeling of
M32 with which we make comparisons.  Our conclusions about the
robustness of those authors' results with regard to size of orbit
library are therefore unaffected by questions of regularization.

\begin{figure}
\figurenum{23}
\epsscale{1.}
\plotone{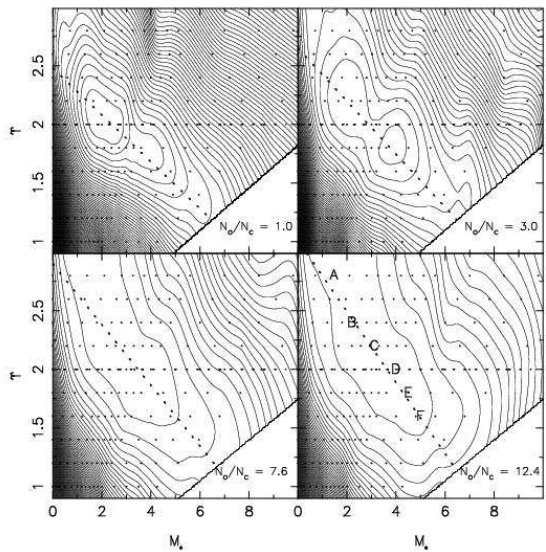}
\label{fig:m32real-four}
\caption{Contour plots of the $\chi^2$ that measures the quality of
the fit to the combined FOS, CFHT and WHT data sets for M32.  These
are the same data fitted by van der Marel (1998) in their 3I modeling
study.  M32 is assumed to be edge-on.  The four panels show the
results using four different sizes of orbit library.  Model parameters
are the black hole mass $\mh$ in $10^6\msun$ and the $V$-band
mass-to-light ratio $\Upsilon$ in solar units.  Dots indicate models
that were calculated.  Labelled positions are models whose fit to the
data is illustrated in detail in Figure~\ref{fig:m32real-kinem}.  The
upper right panel is based on the same number of orbits as in van der
Marel et al. (1998) and shows two distinct $\chi^2$ minima, as in
their paper.  As the number of orbits is increased, these two minima
merge and broaden into a plateau of constant $\chi^2$.}
\end{figure}

Figure~\ref{fig:m32real-four} shows the results of fitting the full
data set using four different orbit numbers, $N_o=(665, 1999, 5127,
8352)$, or $N_o/N_c=(1.0,3.0,7.6,12.4)$.  The top-right-hand panel of
Figure~\ref{fig:m32real-four} was made using almost exactly the same
number of orbits as in vdM98.  This plot exhibits two minima in
$\chi^2(\Upsilon,\mh)$; the corresponding plot in vdM98 (their Fig. 6)
also shows two minima, although displaced slightly from the two in our
Figure~\ref{fig:m32real-four}b.  However as $N_o$ is increased, we
find that the two minima merge into a single, broad plateau.

In their edge-on modeling of M32 from these data, vdM98 selected a
model with $\mh=3.4\times 10^6\msun$ and $\Upsilon \approx 2$ as their
best fit.  That model lies between the two minima seen in the
upper-right panel of Figure~\ref{fig:m32real-four} and somewhere near
the center of the constant-$\chi^2$ plateau in the lower panels.  It
is important to note that the approximate agreement with the results
of vdM98 is a valuable test of our code. The fact that the contours of
vdM98 are not reproduced {\it exactly} are a result of discretization
and differences in the details of the modeling which tend to be more
obvious since the solutions depend on the small numbers of orbits used
($\sim 2000$).

\begin{figure}
\figurenum{24}
\epsscale{0.85}
\plotone{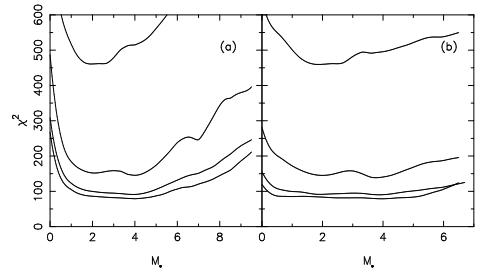}
\label{fig:m32real_1D}
\caption{1-D $\chi^2$ plots along the dotted lines in 
Figure~\ref{fig:m32real-four}. (a) Horizontal cut; (b) slanted
cut.
The four lines in each plot correspond to the four
different numbers of orbits used in the modeling, increasing downward
(cf. Figure~\ref{fig:m32real-four}).
These plots show that the local minima appearing for
small $N_o$ disappear as $N_o$ is increased,
yielding a region of nearly constant $\chi^2$ stretching
at least from $\sim 1\times 10^6\msun$ to $\sim 6\times 10^6\msun$.}
\end{figure}

In Figure~\ref{fig:m32real_1D} we present cuts along two axes in the
$\chi^2(\Upsilon,\mh)$ plots (indicated by the dotted lines in
Figure~\ref{fig:m32real-four}). One cut is at $\Upsilon = 2$ and the
other cut lies roughly along the minimum of the $\chi^2$-valley. These
figures demonstrate that for the largest values of $N_o/N_c$ {\it no
preferred value for $\mh$ in M32 can be found} over a range in values
that extends at least from $\sim 1.5\times 10^6\msun$ to $\sim5\times
10^6\msun$.

\begin{figure*}
\figurenum{25}
\epsscale{1.}
\label{fig:m32real-kinem}
\plotone{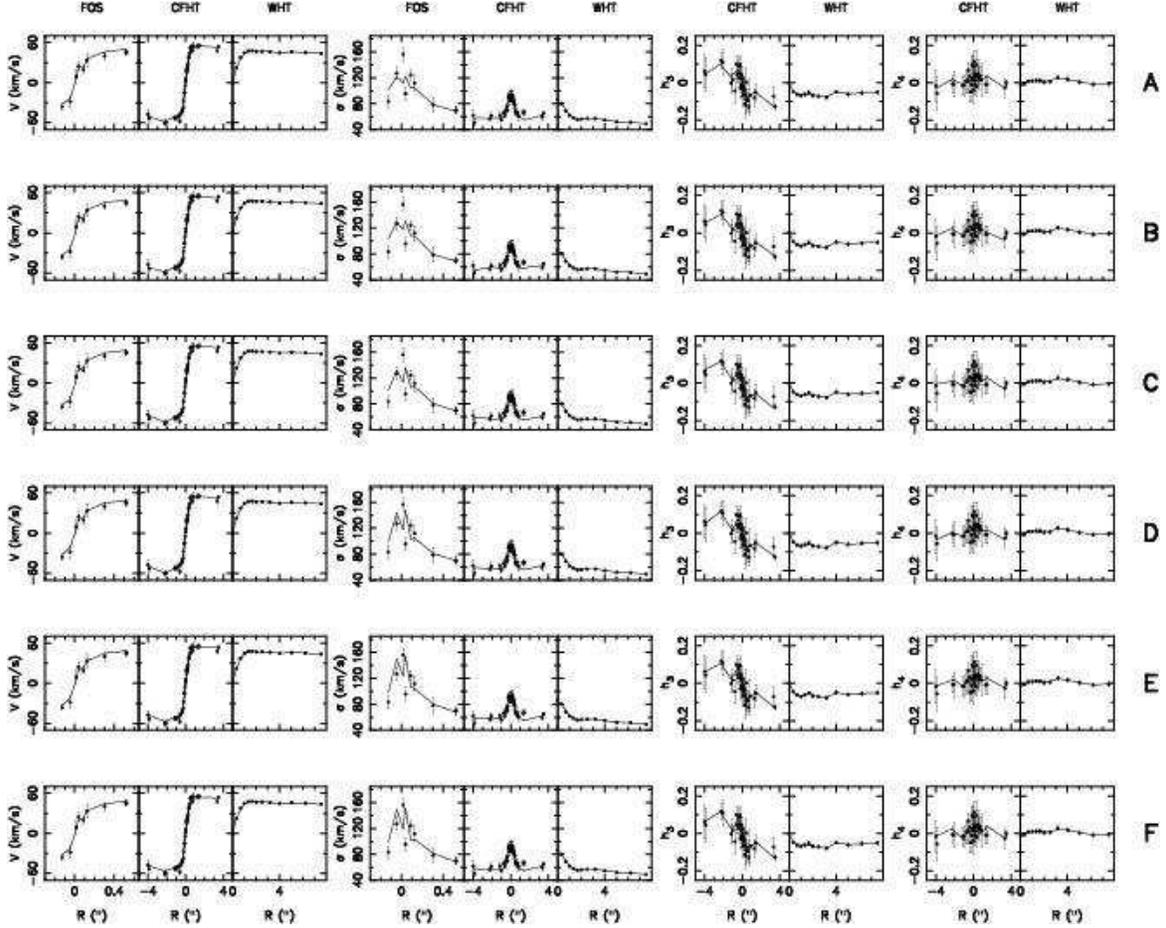}
\caption{Predictions of selected models from
Figure~\ref{fig:m32real-four} compared with a subset of the M32 data.
Models A-F have black hole masses ranging from $1.4\times 10^6\msun$
(Model A) to $4.8\times 10^6\msun$ (Model F).  All plots show fits to
major-axis data; however note that data along other position angles
were also used in constructing the models and the $\chi^2$ values
plotted in Figure~\ref{fig:m32real-four} include the full data.  There
is very little difference in the quality of fit of these models to the
CFHT and WHT data. The only visual difference is between the fits to
the FOS data. The wildly varying velocity dispersions for the FOS data
make this harder to fit. These plots along with Table~1, show that
black holes with masses in the range $1.4\times
10^6\msun<\mh<4.8\times 10^6\msun$ are equally consistent with the
data.}
\end{figure*}

We demonstrate the indeterminacy even more clearly in
Figure~\ref{fig:m32real-kinem}, which shows detailed fits to the
kinematics for a set of models lying along the $\chi^2$ plateau in
Figure~\ref{fig:m32real-four}.  The differences between the various
models -- which span a range of almost a factor of four in $\mh$ --
are almost invisible, with the exception of the predicted values of
$\sigma$ in the FOS apertures.  This could be interpreted to mean that
the FOS data contain useful information about $\mh$, but we consider
this unlikely, since {\it none} of the models fits the FOS data well,
due to the large point-to-point fluctuations in the FOS velocity
dispersions.  It is entirely possible that smoother data, with the
same spatial resolution as the FOS data, could have been fit well by
all the models in this set.  \footnote{Preliminary results of modeling
the M32 STIS data of Joseph et al. (2001) show that these data can be
fit well over a finite range in $\mh$, without the variations apparent
here in the fits to the FOS data (Valluri et al. 2004- in
preparation).}

\begin{table}
\begin{center}
\caption{$\chi^2$ of fit to individual datasets. \label{t:m32real-kinem}}
\bigskip
\begin{tabular}{rrrrrrr}
\\
Model & $\mh$ & $\Upsilon_V$ & $\chi^2_{\rm FOS}$ & $\chi^2_{\rm CFHT}$ &$\chi^2_{\rm WHT}$ & $\chi^2$\\
\\
\hline
 A & 1.4 & 2.8 & 22.45 & 49.24 & 39.55 & 111.19\\
 B & 2.4 & 2.4 & 22.88 & 49.32 & 28.36 & 100.56\\
 C & 3.3 & 2.2 & 25.17 & 51.33 & 26.51 & 103.01\\
 D & 4.0 & 2.0 & 30.96 & 46.97 & 28.11 & 106.04\\
 E & 4.5 & 1.8 & 32.88 & 49.38 & 27.30 & 109.56\\
 F & 4.8 & 1.6 & 33.90 & 50.20 & 25.88 & 109.98\\
\hline
\end{tabular}
\end{center}
\end{table}

In order to make the case for indeterminacy in $\mh$ even more
airtight, we present in Table~\ref{t:m32real-kinem} the contribution
to $\chi^2$ from each of the three partial datasets (FOS, CFHT, WHT)
that make up our data set C. The values of $\chi^2_{\rm CFHT}$ appear
to fluctuate randomly from model A through F with no systematic
behavior.  By contrast, $\chi^2_{\rm FOS}$ is smallest for Models B
and A and increases steadily toward larger $\mh$.  The opposite trend
is observed for the fit to the WHT data; as a result, the total
$\chi^2$ remains almost precisely constant.  (Mass constraints
contribute almost nothing to the total $\chi^2$ since they are fitted
to better than 0.1\% accuracy at all points within the minimum
$\chi^2$-valley.) Although the lower values of $\mh$ provide the
lowest $\chi^2_{\rm FOS}$ they require too large a value of $\Upsilon$
to fit the large radius data.  However the relative difference between
models A and B or F and B is statistically insignificant.

\begin{figure}
\figurenum{26}
\epsscale{0.85}
\label{fig:kin3d_multi}
\plotone{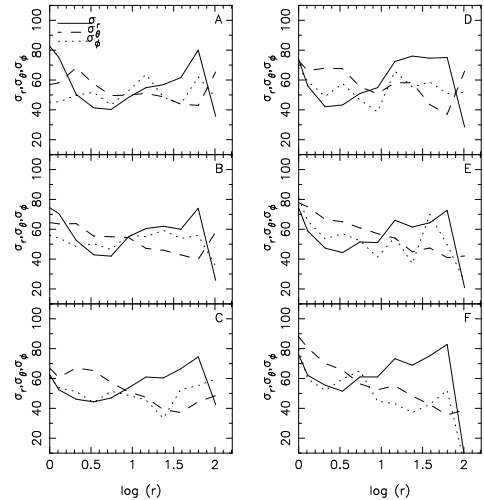}
\caption{ 
Intrinsic 3-D kinematics along the major axis for each of
the models A-F.  Black hole masses ranging from $1.4\times 10^6\msun$
(Model A) to $4.8\times 10^6\msun$ (Model F).  
}
\end{figure}

The internal kinematics of our models must vary with $\mh$ and
$\Upsilon$ in order to maintain fixed observables.
Figure~\ref{fig:kin3d_multi} shows plots of the major axis, internal
velocity dispersion components for models A through F.  The behavior
is precisely as expected: near the center, models with lower $\mh$
maintain a high central velocity dispersion by putting the largest
fraction of stars on radial orbits; at high $\mh$, the central
line-of-sight dispersion is maintained by transferring more and more
stars to nearly circular orbits around the black hole.

\section{The Effect of Adding Regularization Constraints}

In order to test the effect on the solutions of adding smoothness
constraints, we ran a series of models to fit the full set of
kinematical and mass constraints for pseudo data set B, with various
values of the smoothing parameter $\lambda$ in
equation~(\ref{Eq:regularize_chisq}).  In this set of runs, the errors
on the data were selected to be the same as those of the real data.
Since regularization is computationally expensive, we ran this series
of models with only 5000 orbits instead of the full orbit library.

\begin{figure*}
\figurenum{27}
\epsscale{.85}
\label{fig:chi1D_lam_nois}
\plotone{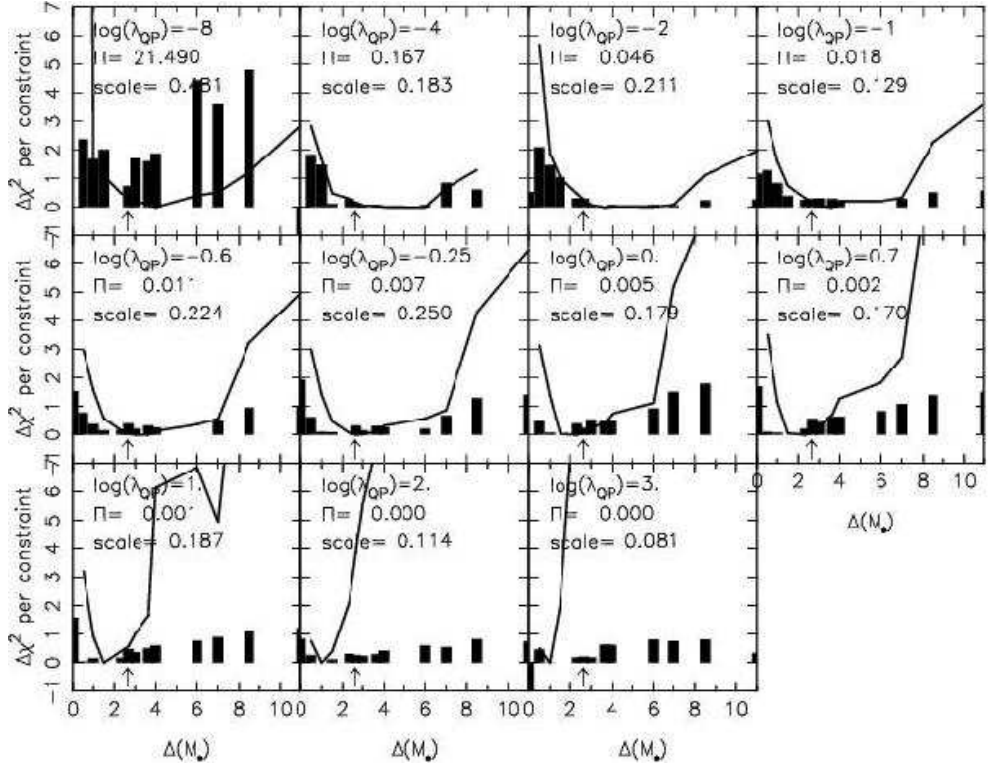}
\caption{ Plots of $\chi^2$ versus $\mh$ and for $\Upsilon_V = 2$ for
a sequence of smoothing parameters ($\lambda$). The quantities $\Pi$
and $scale$ are defined in the text. In each plot the arrow marks the
position of the ``true value of $\mh$''. At each data point the height
of the bar is $\propto (\Pi-\Pi_{\rm min})/(\Pi_{\rm max}-\Pi_{\rm min})$
for the model.}
\end{figure*}

Each plot in Figure~\ref{fig:chi1D_lam_nois} shows $\chi^2$ per
constraint versus $\mh$ for $\Upsilon_V = 2$.  For each choice of
($\mh, \Upsilon_V=2$), the average level $\Pi$ of noise in the
solution was computed via equation~(\ref{eq:noisepi}) in
\S~\ref{sec:regularize} and the value of $\Pi$ is indicated by the
height of the solid bar at each point.  Since $\Pi$ varied by more
than four orders of magnitude as $\lambda$ was varied, the height of
the bar has been rescaled in each plot as $(\Pi_i -
\Pi_{min})/\Pi_{min}$. In addition the plots give the quantities
$scale = (\Pi_{max} - \Pi_{min})/\Pi_{min}$ and $\Pi$, the mean of all
the noise values in a given plot.  Arrows in each plot mark the
position of the true value of $\mh$.

The primary conclusion to be drawn from
Figure~\ref{fig:chi1D_lam_nois} is that adding regularization
constraints does not suddenly or dramatically reduce the degeneracy in
$\mh$. Although the mean level of noise ($\Pi$) drops by a factor of
$\sim 500$ as $\lambda$ increases from $\lambda = 10^{-8}-10^{-2}$,
the flat $\chi^2$ plateau persists over this range with negligible
decrease in the width of the plateau. There is no indication that the
algorithm achieves good fits for incorrect values of $\mh$ by
selecting spuriously noisy solutions.  Indeed, the noise level is
roughly constant along the constant-$\chi^2$ plateaus, and rises
sharply only outside; we provisionally interpret this to mean that all
solutions along the plateau are ``equally good'' and that the
algorithm does not need to construct highly artificial solutions in
order to achieve its good fits.  As $\lambda$ is increased beyond
$\sim 1$, the extended plateau is replaced by a true minimum in
$\chi^2$; this is a necessary consequence of the smoothness
constraint, which begins to penalize solutions characterized by sharp
phase-space gradients, even if they reproduce the data.  However, the
regularization does not seem to have any special ability to select out
the correct $\mh$: instead, as $\lambda$ is increased, the best-fit
$\mh$ systematically drops.  This is also expected, since there is no
reason for the true solution to also be the ``most smooth'' as defined
by any particular choice of penalty function.

\begin{figure}
\figurenum{28}
\epsscale{1.} 
\label{fig:MvsLam}
\plotone{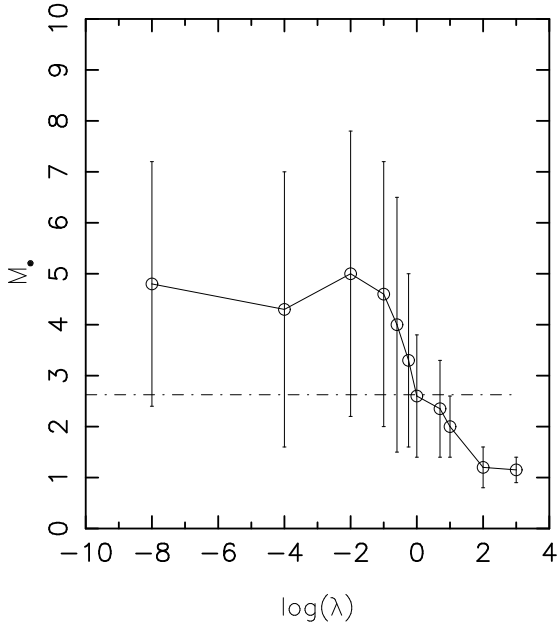}
\caption{ Variation of the mid point of the $\chi^2$ valley as a
 function $\log(\lambda)$. Error bars indicate the range within which
 $\Delta\chi^2 = 0.5$.  The broken line represents the ``true'' value
 of $\mh$ for the pseudo data.}
\end{figure}

Figure~\ref{fig:MvsLam} shows how the best-fit value of $\mh$ and the
range of acceptable $\mh$ values varies with the level of imposed
smoothing.  We defined the range in $\mh$ by $\Delta\chi^2=0.5$.  For
small values of $\lambda$, there is no well-defined minimum in
$\chi^2$ and we defined the best-fit value as the value of $\mh$ at
the center of the $\Delta\chi^2=0.5$ region. 
 While there does exist a particular value of $\lambda$
($\lambda\approx 1$) for which the best-fit $\mh$ is close to the
input value, there would seem to be no way to guess this value based
only on a plot like Figure~\ref{fig:MvsLam}.  When $\log\lambda$ is
increased just slightly above this value, the best-fit $\mh$ drops
below its true value as the smoothness constraint begins to bias the
solution toward overly-smooth phase space distributions.  In other
words, the optimal value of $\lambda$ is only slightly smaller than
the value at which the solutions begin to be seriously biased.

Several authors have based their choice of an optimal smoothing
parameter on the ability of their algorithm to reproduce a specific 2I
distribution function (e.g. Gerhardt et al. 1998, Cretton et al. 1999,
Verolme \& de Zeeuw 2002). There are two potential problems with this
approach: first it has been shown that even for a known distribution
function, the optimal value of $\lambda$ depends on the choice of data
set (e.g. Cretton et al. 1999); second this choice of $\lambda$ is not
guaranteed to give the {\it underlying} distribution function - but
just one of the 3I distribution functions that is close to a 2I form.
This is unlikely to be useful in the general case where the
distribution function could deviate significantly from the 2I form. One
might be particularly concerned about its applicability in modeling
integral field data for galaxies with significant non-2I features:
(counter rotating disks, cores etc). In such cases the use of a
smoothing parameter optimized to recover a 2I distribution function
could artificially restrict the models.

\begin{figure}
\figurenum{29}
\epsscale{1.} 
\label{fig:2Dchisq_nois}
\plotone{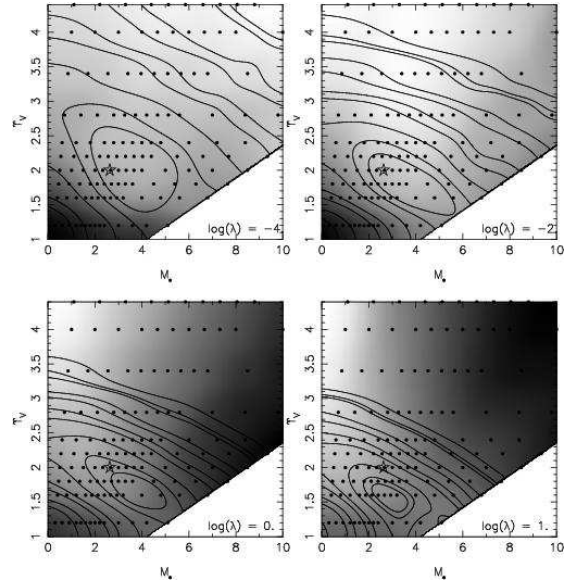}
\caption{2-D $\chi^2$ contour plots for 3 different values of
smoothing parameter $\log(\lambda)$ as indicated. The contours
represent values of $\Delta\chi^2 = \chi^2-\chi^2_{\rm min}$ with the
minimum contour at $\Delta\chi^2 = 0.5$. Subsequent contours are at
intervals of $\Delta\chi^2$ as indicated in the text.  The grey scale
represents the noise. In each plot white represents the least noisy
model and black represents the most noisy model. As before the star
indicates the position of the ``true model''.  }
\end{figure}

In Figure~\ref{fig:2Dchisq_nois} we plot full $\chi^2$ plots for 4
values of $\log(\lambda)$.  The contours trace the value of $\Delta
\chi^2 = (\chi^2-\chi^2_{\rm min})$. The first two contours are at
$\Delta \chi^2 =0.5, 1.$ Subsequent contours are at spacings of
$\Delta \chi^2 = 2.3, 4.6, 6.2, 9.2, 11.8, 18.4$, which are the
68.3\%, 90\%, 95.4\%, 99\% 99.73\% and 99.99\% confidence regions on
$\mh$ and $\Upsilon_V $ jointly \cite{Press96}.  The grey scale
represents the noise. In each plot white represents the least noisy
model and black represents the most noisy model.  As in the case of
the 1-D $\chi^2$ plots it is evident that an elongated $\chi^2$
minimum persists in both parameters even with moderate to high levels
of smoothing.  Once again, noise values do not appear to vary much
within the minimum valley and are comparably low through the entire
$\Delta \chi^2 = 0.5$ contour. There is also no indication that the
models at the extremes of the contours are significantly more noisy
than the models at the center. The regions with the largest amount of
noise are also the regions where the $\chi^2$ values are very
large. Interestingly for models at the top right of the plot, the
$\chi^2$ values are large but the models have little noise. There
appears to be little if any correlation between the noise levels and the
distance from the $\chi^2$ valley.

These experiments are consistent with the view that the potential
estimation problem is inherently ill-conditioned, and that
regularization while it can artificially reduce the solution space can
not overcome the degeneracy.  We note a subtle but important
distinction here between the role of smoothing in 2I and 3I modeling.
In the 2I case, there {\it does} exist a unique (smooth) $f$
corresponding to any assumed potential, and the imposition of
smoothing constraints might be expected to assist in the recovery of
that unique $f$ (\cite{MerrittF96}; \cite{Jalali02};\cite{VerolmedZ02};\\
\cite{CrettonE03}).
Something similar must happen in the 3I case, but as our experiments
show, the additional freedom associated with a 3I $f$ allows many
potentials to be fit with orbital-space populations that are
comparably smooth.  This is just what one expects when solving an
under-determined problem: smoothing alone can not overcome the
degeneracy.  Furthermore, as Figure~\ref{fig:MvsLam} shows, there is a
real danger associated with regularization: if the smoothing parameter
is too large, the best-fit $\mh$ is biased. Indeed this figure
suggests that any $\lambda$ large enough to give a ``best-fit'' $\mh$
-- i.e. a unique minimum in $\chi^2$ -- is also large enough to bias
the location of that minimum.  Hence the conservative approach to
modeling would be to use little or no regularization, even if doing so
means that no ``best-fit'' potential parameters will be found.

Finally, we note here a formal similarity  between the various
sorts of ``constraint'' that can be imposed on $f$.  Forcing $f$ to
depend only on $E$ and $L_z$; restricting the number of orbits from
which a 3I $f$ is constructed; or forcing $f$ to be smooth, via some
regularization scheme, all have the effect of artificially removing
the degeneracy in the potential estimation problem, and converting
flat $\chi^2$ contours into contours with a unique minimum.  We
consider each of these approaches to be dangerous.  While any
combination of $\Phi$ and $f$ that fits the data is acceptable,
statements about the {\it range} of acceptable potentials depend
critically on how flexible the modeling algorithm is at representing
different forms of $f$.  This consideration is particularly important
when attempting to estimate the value of $\mh$ in galactic nuclei,
since artificially restricting $f$ may give the mistaken impression
that a particular value of $\mh$ is preferred, when in a fact a model
with $\mh=0$ can fit the data equally well.

\section{Discussion}

We have shown that the results obtained from stellar dynamical
modeling of galaxy centers can depend as strongly on the flexibility
of the modeling algorithm as on the number and nature of the
observational constraints.  Estimation of the parameters $\mh$ (black
hole mass) and $\Upsilon$ (stellar mass to light ratio) that define
the gravitational potential is typically an under-determined
(degenerate) problem, and 3I (three-integral) 
modeling can (and, we believe, often
does) generate spurious, ``best-fit'' model parameters that bear no
special relation to the true parameters.  We demonstrated this in the
case of previously-modeled data for M32: increasing the number of
orbits by a factor of $\sim 4$ above what was used in the earlier
studies led us to substantially different conclusions about the most
likely value of the black hole mass and its uncertainty in this
galaxy.  Indeed, we found that no single value of $\mh$ was preferred
and that values for $\mh$ in the range $1.5\times 10^6\msun$ to
$5\times 10^6\msun$ could reproduce the data with no appreciable
change in the goodness of fit
(Figures~\ref{fig:m32real-four}-\ref{fig:m32real-kinem}).  We argued
that the degeneracy is not a numerical artifact, but derives instead
from the great freedom available to fit a given, oblate-spheroidal
mass distribution using a 3I distribution function.

Our work raises three questions about published and ongoing modeling
studies of galactic nuclei.  

\begin{enumerate}

\item{Does a given data set contain enough information to distinguish
a best-fit value of $\mh$, or is $\mh$ indeterminate, and if so, over
what range of values?}

\item{How can one be certain that a given modeling algorithm 
accurately reproduces the true interval of  $(\mh,\Upsilon)$
values allowed by the data?}

\item{What is the role of regularization (via maximum entropy or any
other scheme) in reducing the solution space?}

\end{enumerate}

With regard to the first question, we showed that the degeneracy in
$\mh$ is substantial even for one of the best available data sets, the
pre-STIS data for M32 (vdM98).  These data resolve
the sphere of influence of the black hole (${\rm FWHM}/2r_h\approx
0.25$ assuming $\mh\approx 3\times 10^6\msun$); include Gauss-Hermite
moments up to $h_6$; extend outward to $\sim 10r_h$ along several
position angles; and have a high signal to noise ratio.  Furthermore,
by virtue of its high rotation, M32 allows tighter constraints to be
placed on the orbital distribution and on $\mh$ than in ``hotter''
stellar systems.  Nevertheless our constraints on $\mh$ were weak,
spanning a factor of $\sim 3.5$.

We expect the degree of degeneracy in quantities like $\mh$ to depend
on the quality of the data, in particular on the degree to which the
black hole's sphere of influence is resolved.  We demonstrated this
explicitly in \S5-6 using our simulated and real data sets: placing
useful constraints on $\mh$ requires an effective resolution ${\rm
FWHM}/2r_h$ that is substantially less than one.  Another relevant
factor is the radial extent of the data, which determines how well the
mass to light ratio is constrained (cf. Figure~\ref{fig:stswht}).

With regard to the second question, we showed that $\chi^2$ contours
for the simulated and true M32 data sets change significantly when the
number of orbits used increased from twice to in excess of $10$
times the total number of data points (kinematical plus mass
constraints).  Fully general statements about the minimum number of
orbits required to explore the full extent of the allowed solution
space are impossible to make, since some data points are clearly more
useful than others for constraining quantities like $\mh$.  For
instance, we argued (\S~5) that direct fitting to LOSVDs is less
efficient than fitting to Gauss-Hermite moments, in the sense that
more orbits are required in the former case to achieve the same degree
of modeling flexibility.  In one data set treated above (the simulated
WHT data, Figures~\ref{fig:vdm3_NobyNc} and \ref{fig:vdm3_NobyNc1d}),
the shape of the $\chi^2$ contours continued to change as the ratio of
orbits to constraints was increased from $\sim 10$ to $\sim 20$ and it
is conceivable that even more orbits would be required to reproduce
the true extent of the indeterminacy in $\mh$.  A number of different
factors are likely to influence the minimum number of orbits required
to constrain $\mh$ and the distribution function of the surrounding spheroid: the type of
galaxy, the quality, nature and dimensionality of the data (spatial
and spectral resolution, single slit, multiple slits, 2D spatial
coverage), the true internal kinematics of the galaxy, etc.

The importance of 2D coverage to constrain 3I distribution functions
has been recently illustrated by modeling studies based on data from integral
field spectrographs such as SAURON on the WHT.  Verolme et al. (2002)
presented 3I modeling of M32 based on data from the SAURON as well as
HST/STIS data from Joseph et al. (2001). It is likely that 2D data
such as those presented by Verolme et al. (2002) are able to greatly
reduce the degeneracy which we demonstrated in single- or multiple
slit data sets. They showed for instance that 2D data are significantly
better at constraining the mass-to-light ratio $\Upsilon$, than are
multiple long slits.  However the number of data constraints modeled
by them was $\sim 8000$ and the number of orbits used was only $1960$,
giving $N_o/N_c\lesssim 0.25$. Mathematically a unique solution will
always be found if $N_o<N_c$. The well-defined minima in their
$\chi^2(\Upsilon,\mh)$ plots could be a consequence of the smaller
ratio of orbits to constraints, or could mean that their data have
overcome the degeneracy. Testing the latter hypothesis will be
difficult however given the large number of orbits ($N_o\gtrsim 10^5$)
that would be required.

As pointed out in \S~\ref{sec:indeterm} it has not been demonstrated
mathematically that it is possible to construct a unique 3I distribution function from
projected kinematical, surface brightness data no matter how perfectly
the LOSVDs are sampled. It is often stated (e.g. Cappellari et
al. 2003b) that 2D kinematical coverage is essential to constrain the
orbital structure in a galaxy from observables. Thus it would appear
that limited data (e.g. slits along one or more axes) are guaranteed
to be insufficient. Most published estimates of black hole masses are
based on multiple slit data and are therefore likely to suffer from
this indeterminacy.

If the potential estimation problem is generically under-determined,
why has the degeneracy escaped general notice until now? 
In fact the degeneracy is apparent in a number of published 
modelling studies.  
Two examples are
the Gebhardt et al. (2000a) study of NGC~3379 and the Cretton \& van
den Bosch (1999) study of NGC~4342.  In the former study, the modeling
used 6400 orbits compared with 702 kinematical constraints and 100
mass constraints, or $N_o/N_c=8.0$.  Goodness-of-fit contours
generated from 3I models show a plateau of nearly-constant $\chi^2$
extending from $\sim 10^6\msun$ to at least $\sim 10^8\msun\ $ (their
Fig. 7).  In fact the authors state that ``the difference between the
no-black hole and black hole models is so subtle that one can barely
discriminate those models'' (cf. their Fig. 11).  Gebhardt et
al. nevertheless argue for an (inclination-dependent) best-fit value
of $\mh$ based on the (puzzlingly asymmetric) wings of the central
LOSVD as derived from FOS data.  In the Cretton \& van den Bosch
(1999) study of NGC 4342, the authors again found that a model with no
black hole provides ``fits to the actual data [that] look almost
indistinguishable'' from that of a model with $\mh=3.6\times 10^8
\msun\ $, the claimed best-fit value (cf. their Fig. 8).  This study
used 1400 orbits compared with $\sim 250$ constraints, or $N_o/N_c
\approx 5.6$.  We note that both of these data sets have ${\rm
FWHM}/2r_h\approx 0.5$ (if $\mh$ is computed from the \msig\
relation), consistent, according to our analysis, with the fact that
no best-fit value of $\mh$ could be found.

Since about 1999, it has been common practice in 3I modeling to
include smoothness constraints on the orbital weights, in the form of
maximum entropy or some other regularization scheme (e.g. Cretton et
al. 1999, Gebhardt et al. 2000a; Verolme et al. 2002). We showed above
that imposing smoothness constraints has effects similar to those of
other, ad hoc restrictions on the form of $f$: they reduce the
flexibility of the modeling algorithm to adjust $f$ in response to
changes in $\Phi$, and therefore tend to ``select out'' a particular
$\Phi$ as preferred.  When the smoothing is kept at a level too low to
bias $f$, the $\chi^2$ contours on $\mh$ show the correct,
perfectly-flat form associated with ill-conditioned estimation
problems.  We believe that it would be appropriate to repeat a number
of the published modeling studies, giving careful attention to the
role of regularization on the range of allowed solutions.

Standard practices for estimating and describing confidence intervals
will need to be changed when dealing with indeterminate problems like
the estimation of $\mh$ in galactic nuclei.  Quoting a black hole mass
as $5.0\pm 2 \times 10^8\msun$, for instance, is inappropriate if there
is no best-fit value.  Instead, a notation like $\mh=(3-7)\times
10^8\msun$ would more correctly convey the result that any value in
the specified range is equally likely.  In addition, when using
estimated black hole masses as data points in other statistical
studies, care will have to be taken to deal correctly with the
degeneracy.  For instance, standard least-squares fitting assumes that
there exists a best estimate of the measured quantities and that the
errors about that estimate are normally distributed.  Both assumptions
are incorrect when the measured quantity is indeterminate.

An important motivation for measuring black hole masses in galactic
nuclei is to refine and extend the \msig\ relation.  Past discussions
of the uncertainties in that relation have focused on statistical
techniques \cite{MerrittF01a} or on systematic differences in the
definitions of $\sigma$ \cite{Tremaine02}.  We suggest that the
largest source of uncertainty in the \msig\ relation is likely to be
the degeneracy in SBH masses as determined from stellar kinematical
data.

\section{Conclusions}

\begin{enumerate}

\item{ The axisymmetric potential estimation problem is generically
under-determined: a range of values for quantities like $\mh$, the
black hole mass, and $\Upsilon$, the mass-to-light ratio of the stars,
can generally be found that are equally consistent with the observed
kinematics.  The indeterminacy arises from the large number of
distinct distribution functions $f$ that can reproduce a given mass
model.}

\item{The indeterminacy becomes apparent only when the modeling
algorithm is flexible enough to represent a wide range of stellar
distribution functions.  In practice, this means having a sufficient
number of distinct orbits or phase-space cells.  When the orbit
library is too small, spurious minima appear in plots like
$\chi^2(\mh)$ due to the algorithm's inability to reproduce certain
orbital populations as well as others.}

\item{When the LOSVDs are well sampled, there is no advantage to
fitting the full LOSVD over fitting just the GH moments, even when
they have large wings. The only exceptions are likely to be when
LOSVDs are multimodal}.

\item{A re-analysis of data for M32 published prior to 2000 reveals
that these data do not imply a preferred or best-fit value for the
black hole mass, contrary to claims made in the literature.  We show
that a range of values, $1.5\times 10^6\msun<\mh<5\times 10^6\msun$, are
equally consistent with these data.  We demonstrate that the best-fit
values of $\mh$ in M32 derived in earlier studies are likely to have been
biased by the use of too few orbits to represent $f$.}

\item{Regularization reduces the range of acceptable models, but we
find no indication that the true potential can be recovered simply by
enforcing smoothness.  For a given smoothing level, all solutions in
the minimum-$\chi^2$ valley exhibit similar levels of noise; as the
smoothing is increased, there is a systematic shift in the midpoint of
the $\chi^2$ valley, until at a high level of smoothing the solution
is biased with respect to the true solution.}

\end{enumerate}

\acknowledgements

We thank J. Magorrian for the use of his multipole-expansion routines.
We thank R. van der Marel for providing us with the M32 data, for his
help with many aspects of the development of the code and for
providing us with the PSF convolution and FFT routines used in the
code. MV would like to thank P.T. de Zeeuw, C. Joseph and
P. Vandervoort for important discussions during various phases of this
work. Last but not the least we thank the referee Ortwin Gerhard for
his detailed comments and for urging us to include a discussion on the
effect of regularization constraints in this paper. This work was
supported by NSF grants AST 96-17088 and AST 00-71099 and by NASA
grants NAG5-6037 and NAG5-9046 and STScI grant HST-AR-08759.




\begin{thebibliography}{}
\bibitem[Barth et al. 2001]{Barth01}
	Barth, A. J. et al. 2001,
	ApJ, 555, 685
\bibitem[Bender, Kormendy \& Dehnen 1996]{Bender96} 
	Bender, R., Kormendy, J.\& Dehnen, W. 1996, 
	ApJ, 464, L123
\bibitem[Binney, Davies \& Illingworth 1990]{BinneyDI90}
	Binney, J. J., Davies, R. L. \& Illingworth, G. D. 1990,
	ApJ, 361, 78
\bibitem[Bower et al. 2001]{Bower01}
	Bower, G. A. et al. 2001,
	ApJ, 550, 75
\bibitem[Cappellari et al. 2003a]{Cappellari02}
	Cappellari, M. et al. 2003, 
	ApJ 578, 787
\bibitem[Cappellari et al. 2003b]{Cappellari03}
	Cappellari, M. et al. 2003,in Carnegie Observatories Astrophysics 
        Series, Vol. 1: Coevolution of Black Holes and Galaxies, 
        ed. L. C. Ho (Pasadena: Carnegie Observatories)(astro-ph/0302274).
\bibitem[Contopoulos 1960]{Contopoulos60}
         Contopoulos, G. 1960, Zeitsch. Astrop., 49, 273
\bibitem[Cretton \& van den Bosch 1999]{CrettonB99}
	Cretton, N. \& van den Bosch, F. C. 1999,
	ApJ, 514, 704
\bibitem[Cretton et al. 1999]{Cretton99}
	Cretton, N., de Zeeuw, P. T., van der Marel, R. P., \&
	 Rix, H.-W. 1999, 
	ApJS, 124, 383
\bibitem[Cretton \& Emsellem 2003]{CrettonE03}
         Cretton, N. \& Emsellem, E. 2003, MNRAS, Submitted.
\bibitem[Dehnen 1995] {Dehnen95} Dehnen, W. 1995, MNRAS, 274, 919
\bibitem[Dejonghe 1986]{Dejonghe86}
	Dejonghe, H. 1986,
	Phys. Rep. 133, 218
\bibitem[Dejonghe \& Merritt 1992]{DejongheM92}
	Dejonghe, H. \& Merritt, D. 1992, 
	ApJ, 391, 531
\bibitem[Emsellem, Monnet \& Bacon 1994]{EmsellemMB94}
	Emsellem, E., Monnet, G., \& Bacon, R. 1994, 
	A\& A, 285, 723
\bibitem[Emsellem, Dejonghe \& Bacon 1999]{Emsellem99}
	Emsellem, E., Dejonghe, H. \& Bacon, R. 1999,
	MNRAS, 303, 495
\bibitem[Ferrarese 2002]{Ferrarese01}
	Ferrarese, L. 2002,
	in Current High-Energy Emission around Black Holes,
	proceedings of the 2nd KIAS Astrophysics Workshop,
        ed. C.-H. Lee. (Singapore: World Scientific), p3.
	(astro-ph/0203047)
\bibitem[Ferrarese and Merritt 2000]{FerrareseM00}
	Ferrarese, L. and Merritt, D. 2000,
	ApJ, 539, L9
\bibitem[Gebhardt et al. 2000a]{Gebhardt00a}
	Gebhardt, K. et al. 2000a,
	AJ, 119, 1157
\bibitem[Gebhardt et al. 2000b]{Gebhardt00b}
	Gebhardt, K. et al. 2000b,
	ApJ, 539, L13
\bibitem[Gebhardt et al. 2003]{Gebhardt02}
	Gebhardt, K. et al. 2003,
        ApJ 583, 92
\bibitem[Genzel et al. 1997]{Genzel97}
	Genzel, R., Eckart, A., Ott, T., and Eisenhauer, F. 1997,
	MNRAS, 291, 219
\bibitem[Gerhard 1993]{Gerhard93} 
	Gerhard, O. E. 1993, 
	MNRAS, 265, 213
\bibitem[Gerhard \& Binney 1996]{GerhardB96}
	Gerhard, O. W. \& Binney, J. J. 1996,
	MNRAS, 279, 993
\bibitem[Gerhard et al. 1998]{Gerhard98}
        Gerhard, O, E., Jeske, G., Saglia, R.P. \& Bender, R.118, MNRAS, 
        295, 197
\bibitem[Ghez et al. 1998]{Ghez98} 
	Ghez, A. M., Klein, B. L., Morris, M., and Becklin, E. E. 1998,
	ApJ, 509, 678
\bibitem[Hairer \& Wanner (1993)]{Hairer93}
	Hairer, E. \& Wanner, G. 1993,
	Solving Ordinary Differential Equations II. 
        Stiff and Differential-Algebraic Problems,
        Springer Series in Comput. Mathematics, Vol. 14
	(Springer-Verlag: Berlin)
\bibitem[Hughes et al. 2001]{Hughes01}
	Hughes, M., Axon, D. J., Alonso-Herrero, Almudena, and
	Atkinson, J. 2001, in
	The Central Kiloparsec of Starbursts and AGN:
	The La Palma Connection,
	Astron. Soc. Pac. Conf. Ser. Vol. 249,
	ed. J. H. Knapen, J. E. Beckman, I. Shlosman \& T. J. Mahoney
	(ASP: Chelsea, Michigan), 363
\bibitem[Hunter 1975]{Hunter75}
	Hunter, C. 1975,
	AJ, 80, 783
\bibitem[Hunter \& Qian 1993]{HunterQ93}
	Hunter, C. \& Qian, E. E. 1993,
	MNRAS, 262, 401
\bibitem[Jalali \& de Zeeuw 2002]{Jalali02}
        Jalali, M.~A.~\& de Zeeuw, P.~T.\ 2002,
        \mnras, 335, 928
\bibitem[Joseph et al. 2001]{Joseph01}
	Joseph, C. L. et al. 2001,
	ApJ, 550, 668
\bibitem[Kulessa \& Lynden-Bell 1992]{Kulessa92}
	Kulessa, A. S. \& Lynden-Bell, D. 1992,
	MNRAS, 255, 105
\bibitem[Lawson \& Hanson 1995]{Lawson95}
        Lawson, C. L. \& Hanson, R. J. 1995, Solving Least Squares Problems,
        (SIAM: Philadelphia, PA), 269
\bibitem[Little \& Tremaine 1987]{Little87}
	Little, B. \& Tremaine, S. 1987,
	ApJ, 320, 493
\bibitem[Lynden-Bell 1962]{LyndenBell62}
	Lynden-Bell, D. 1962,
	MNRAS, 123, 447
\bibitem[Lynden-Bell 1969]{Lynden-Bell69}
	Lynden-Bell, D. 1969,
	Nature, 223, 690
\bibitem[Magorrian et al. 1998]{Magorrian98}
	Magorrian, J. et al. 1998,
	AJ, 115, 2285
\bibitem[Merritt 1993a]{Merritt93a} 
	Merritt, D. 1993a, 
	ApJ, 413, 79 
\bibitem[Merritt 1993b]{Merritt93b} 
	Merritt, D. 1993b, in
	Structure, Dynamics and Chemical Evolution of Elliptical Galaxies,
	ed. I. J. Danziger, W. W. Zeilinger and K. Kj\"ar
	ESO Conference and Workshop Proceedings No. 45
	(Munich: ESO), p. 275
\bibitem[Merritt 1997]{Merritt97}
	Merritt, D. 1997,
	AJ, 114, 228
\bibitem[Merritt \& Ferrarese 2001a]{MerrittF01a}
	Merritt, D. \& Ferrarese, L. 2001a,
	ApJ, 547, 140
\bibitem[Merritt \& Ferrarese 2001b]{MerrittF01b}
	Merritt, D. \& Ferrarese, L. 2001b, in
	The Central Kiloparsec of Starbursts and AGN:
	The La Palma Connection,
	Astron. Soc. Pac. Conf. Ser. Vol. 249,
	ed. J. H. Knapen, J. E. Beckman, I. Shlosman \& T. J. Mahoney
	(ASP: Chelsea, Michigan), 335
\bibitem[Merritt, Ferrarese \& Joseph 2001]{MerrittFJ01}
	Merritt, D., Ferrarese, L. \& Joseph, C. L. 2001, 
        Science, 293, 1116
\bibitem[Merritt \& Fridman 1996]{MerrittF96}
         Merritt, D. \& Fridman, T. 
         ApJ, 460, 136
\bibitem[Merritt, Meylan \& Mayor 1997]{MerritM97}
	Merritt, D., Meylan, G. \& Mayor, M. 1997,
	AJ, 114, 1074
\bibitem[Merritt \& Saha 1993]{MerrittS93}
	Merritt, D. \& Saha, P. 1993,
	ApJ, 409, 75
\bibitem[Merritt \& Tremblay 1993]{MerrittT93}
	Merritt, D. \& Tremblay, B. 1993,
	AJ, 106, 2229
\bibitem[Merritt \& Tremblay 1994]{MerrittT94}
	Merritt, D. \& Tremblay, B. 1994,
	AJ, 108, 514
\bibitem[Miller 1974]{Miller74}
        Miller, G.F. 1974 in Numerical Solutions of Integral Equations,
        ed. L.M. Delves \& J. Walsh (Oxford: Claredon Press), 175
\bibitem[Miyoshi et al. 1995]{Miyoshi95}
	Miyoshi, M., Moran, J., Herrnstein, J., Greenhill, L., 
	Nakai, N., Diamond, P., and Inoue, M. 1995,
	Nature, 373, 127
\bibitem[Monnet, Bacon \& Emsellem 1992]{Monnet92}
	Monnet, G., Bacon, R., \& Emsellem, E. 1992,
	A\& A, 253, 366
\bibitem[Ollongren 1962]{Ollongren62}
	Ollongren, A. 1962,
	Bulletin of the Astronomical Institute of the Netherlands, 16, 241
\bibitem[Phillips 1962]{Phillips62}
        Phillips, D. L. 1962, J. Ass. Comput. Mach. 9, 84
\bibitem[Press et al. 1992]{Press96}
        Press, W.H., Teukolsky, S.A., Vetterling, W.T. \& Flannery, B.F. 1992,
        Numerical Recipes (Cambridge: Cambridge University Press)
\bibitem[Qian et al. 1995]{Qianetal95} Qian, E. E., de Zeeuw, P. T., van 
	der Marel, R. P. \& Hunter, C. 1995, MNRAS, 274, 602
\bibitem[Richstone 1982]{Richstone82}
	Richstone, D. O. 1982,
	ApJ, 252, 496
\bibitem[Richstone \& Tremaine 1988]{Richstone88}
        Richstone, D.O, \& Tremaine, S. 1988, ApJ, 327, 82
\bibitem[Richstone et al. 1998]{Richstone98}
	Richstone, D. O. et al. 1998, 
	Nature, 395, A14
\bibitem[Rix et al. 1997]{Rix97}
	Rix, H. W., de Zeeuw, P. T., Cretton, N., van der Marel, R. P., \& 
	Carollo, C. M. 1997,
	ApJ, 488, 702
\bibitem[Romanowsky \& Kochanek 1997]{RomanowskyK97}
	Romanowsky, A. J. \& Kochanek, C. S. 1997,
	MNRAS, 287, 35
\bibitem[Rybicki 1987]{Rybicki87}
	Rybicki, G. B. 1987,
	in Structure and Dynamics of Elliptical Galaxies,
	IAU Symposium No. 127, ed. T. de Zeeuw
	(Dordrecht: Reidel), 397
\bibitem[Sarzi et al. 2001]{Sarzi01}
	Sarzi, M. et al. 2001,
	ApJ, 550, 65
\bibitem[The \& White 1986]{The86}
	The, L. S. \& White, S. D. M. 1986,
	ApJ, 92,1248
\bibitem[Tikhonov 1963]{Thikonov63}
        Tikhonov, A. N. 1963, Soviet Math. 4, 1035
\bibitem[Tremaine et al. 2002]{Tremaine02}
	Tremaine, S. et al. 2002,
	ApJ 574, 740
\bibitem[van Albada \& van Gorkom 1977]{vanAlbada77}
	van Albada, T. S. \& van Gorkom, J. H. 1977,
	A\& A, 54, 121
\bibitem[van der Marel 1999]{vanderMarel99}
	van der Marel, R. P. 1999,
	in IAU Symp. 186, Galaxy Interactions at Low and High Redshift
	(Dordrecht: Kluwer), 333
\bibitem[van der Marel et al. 1994a]{vanderMarel94a} 
	van der Marel, R. P., Rix, H.-W., Carter, D., Franx, M., 
	White, S. D. M. \& de Zeeuw, T. 1994a, 
	MNRAS, 268, 521
\bibitem[van der Marel et al. 1994b]{vanderMarel94b} 
	van der Marel, R. P., Wyn-Evans, N., Rix, H.-W., White, S. D. M. 
	\& de Zeeuw, T. 1994, 
	MNRAS, 271, 99 
\bibitem[van der Marel, de Zeeuw \& Rix 1997]{vanderMarel97} 
	van der Marel, R. P., de Zeeuw, P. T. \& Rix, H. W. 1997, 
	ApJ, 488, 119
\bibitem[van der Marel et al. 1998]{vanderMarel98} van der Marel, R. P., 
	Cretton, N., de Zeeuw, P. T. \& Rix, H. W. 1998,
        ApJ, 493, 613 (vdM98)
\bibitem[Verolme \& de Zeeuw 2002]{VerolmedZ02}
         Verolme \& de Zeeuw 2002,
         MNRAS 331, 959
\bibitem[Verolme et al. 2002]{Verolme02}
	Verolme, E. K. et al. 2002,
	MNRAS, 335, 517
\end{thebibliography}
\end{document}